\documentclass[lettersize,journal]{IEEEtran}
\usepackage{amsmath,amsfonts}
\usepackage{algorithmic}
\usepackage{algorithm}
\usepackage{array}
\usepackage[caption=false,font=normalsize,labelfont=sf,textfont=sf]{subfig}
\usepackage{textcomp}
\usepackage{stfloats}
\usepackage{url}
\usepackage{verbatim}
\usepackage{graphicx}
\usepackage{cite}
\usepackage{lineno}
\usepackage{dsfont}
\newcommand{\D}{\textnormal{d}}
\begin{document}

\title{Impact of coherent mode coupling on noise performance in elliptical aperture VCSELs for datacom}

\author{Cristina Rimoldi, Lorenzo L. Columbo, Alberto Tibaldi,~\IEEEmembership{Member,~IEEE,} Pierluigi Debernardi, Sebastian Romero Garc\'ia, Christian Raabe, Mariangela Gioannini,~\IEEEmembership{Member,~IEEE}
        % <-this % stops a space
\thanks{Cristina Rimoldi, Lorenzo L. Columbo, Alberto Tibaldi, and Mariangela Gioannini are with the
Dipartimento di Elettronica e Telecomunicazioni, Politecnico di Torino, corso Duca degli Abruzzi 24, Torino, IT-10129
Italy (e-mail: cristina.rimoldi@polito.it; lorenzo.columbo@polito.it; alberto.tibaldi@polito.it;
mariangela.gioannini@polito.it). Pierluigi Debernardi is with Consiglio Nazionale delle Ricerche (CNR), Istituto di Elettronica e di Ingegneria dell’Informazione e delle Telecomunicazioni (IEIIT), corso Duca degli Abruzzi 24, Torino, IT-10129, Italy (e-mail:pierluigi.debernardi@ieiit.cnr.it). Sebastian Romero Garc\'ia and Christian Raabe are with Cisco Optical, Nordostpark 12, Nuremberg, D-90411, Germany.}% <-this % stops a space
\thanks{Manuscript received XXX NN, YYYY.}}

\markboth{IEEE Journal of Selected Topics in Quantum Electronics,~Vol.~XX, No.~X, XXX~YYYY}%
{Shell \MakeLowercase{\textit{et al.}}: A Sample Article Using IEEEtran.cls for IEEE Journals}

\IEEEpubid{0000--0000/00\$00.00~\copyright~2024 IEEE}

\maketitle

\begin{abstract}
We study the dynamical behavior of medium-size multimode VCSELs with an elliptical oxide aperture, selected for the best trade-off between high output power and modulation speed for datacom applications, with a focus on their relative intensity noise (RIN) performance. Our experimental results, collected for various VCSELs, outline the presence of several peaks in the RIN spectra within the bandwidth of the transmission system, which can limit the eye opening under direct current modulation. Here, we present a rigorous model to explain for the first time the origin of these peaks. In particular, the frequencies of the spectral RIN peaks are analytically described as the result of the non-trivial interaction among transverse modes by addressing the laser dynamics and the related noise features through a time-domain mode expansion approach, accounting for coherent effects in multimode competition, spatial hole burning, and carrier diffusion. 
The laser modulation performance is addressed through dynamical simulations with PAM2 and PAM4 modulations, which clearly demonstrate the potential for high-bitrate optical interconnects. Finally, we address the effect of the oxide aperture aspect ratio via electromagnetic simulations, demonstrating how the ellipticity affects the modal frequency detuning and the RIN, thus providing design guidelines for VCSELs with low RIN performance and outlining a clear roadmap for a substantially improved bandwidth-power trade-off in these devices.
\end{abstract}


\section{Introduction}
\begin{figure*}[t]
\centering
\includegraphics[width=\textwidth]{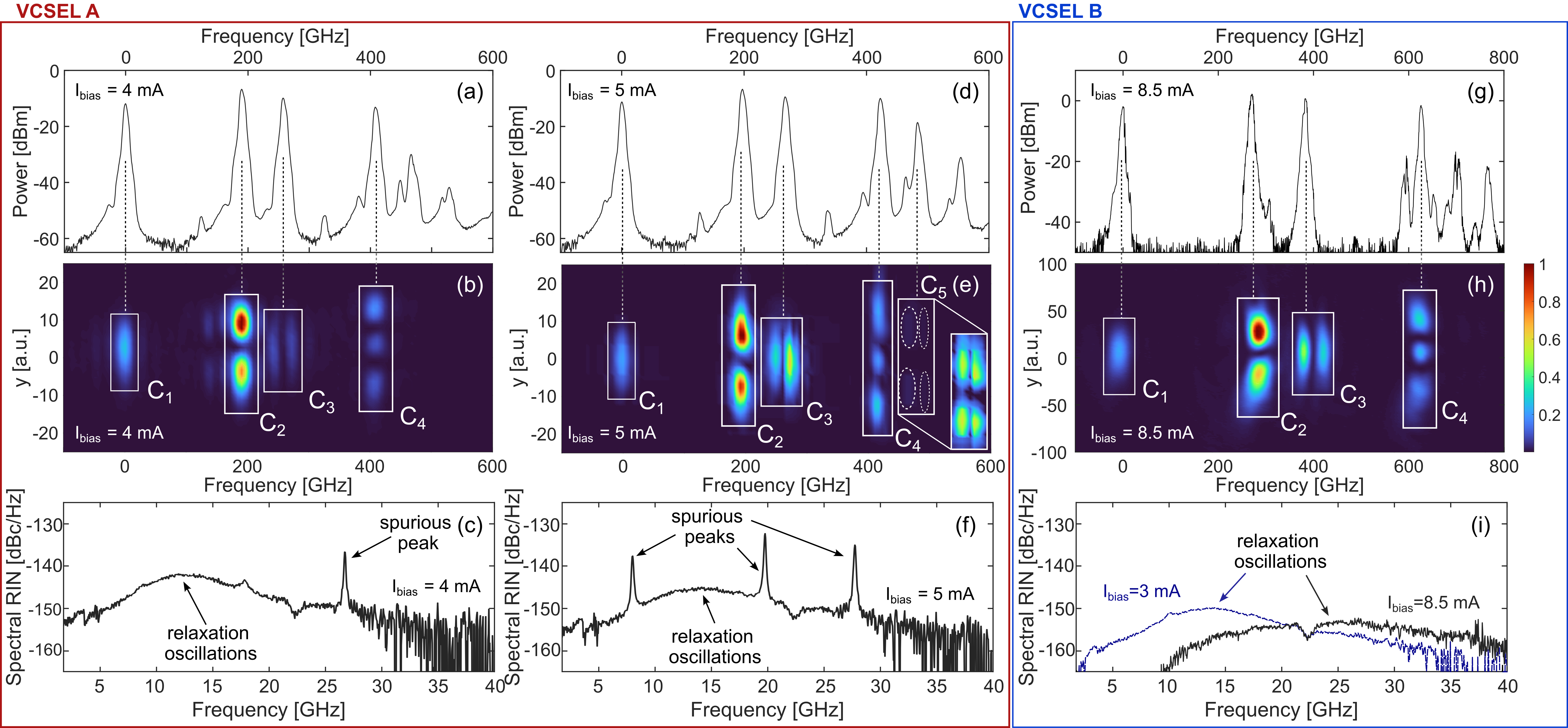}
\caption{VCSEL A: optical spectra (a,d), near-field measurement (b,e), and spectral RIN (c,f) at $I_{bias}=4$ (a-c) and 5\,mA (d-f). VCSEL B: Optical spectrum (g), near-field measurement (h), and spectral RIN (i) at $I_{bias}=8.5$\,mA. The x-axis in (a,d,g) displays the frequency detuning with respect to the fundamental mode. In (b,e,h) this axis has been linearly scaled, based on (a,d,g) respectively, from spatial to frequency units, given the proportionality resulting from the use of diffraction gratings.}
\label{fig:exp}
\end{figure*}
Short-reach data transmission multimode fiber links often make use of multimode 850 nm VCSELs because they display the best trade-off between cost effectiveness, low-threshold current, high power and high-modulation speed as required for intra-data center low-cost communication \cite{Bhatt2024}. Standard circular aperture configurations at the required powers lead to almost degenerate transverse higher order modes, with emission wavelengths separated by a few GHz. While these modes are orthogonal to each other, their beating (mediated by spatial hole burning in the quantum wells) is different from zero and results in undesired peaks in the relative intensity noise (RIN) spectrum with consequent increase of the RIN integrated over the receiver bandwidth, which is detrimental for error-free transmission. In this context, a design of particular interest relies on an elliptical oxide aperture \cite{Gazula2019,Wang2019,Debernardi2002}: by breaking the quasi-degeneracy of the transverse modes, this configuration pushes their frequency separation beyond the receiver bandwidth while also allowing for high emitted power. However, as the required bandwidth will increased due to the demand for increasingly higher data rates, the problem of mastering the RIN characteristics and finding a methodology for optimizing the VCSEL design with low RIN over a wide bandwidth is bound to re-emerge. This work is dedicated to understanding the electromagnetic, static, and dynamical performance of multimode VCSELs with elliptical oxide aperture in order to better address this issue. Based on experimental results, acquired for two different VCSELs with elliptical oxide aperture, we developed a novel and unique numerical tool describing the laser dynamics that accounts for the coherent coupling of transverse modes in the gain medium.\\ 
\IEEEpubidadjcol
The VCSELs have been designed so that one polarization was suppressed, leading to stable emission on one polarization axis and allowing to avoid any relevant polarization switching issues.
In the past decades, coherent longitudinal mode coupling has often been studied in multi-longitudinal edge-emitting lasers (e.g., in \cite{Lenstra2014}) in order to explain mode partition noise in Quantum Well (QW) lasers \cite{Furfaro2004}, the extension of modulation bandwidth via photon-photon resonance \cite{Chacinski2010}, and the self-generation of optical frequency combs in Fabry-Perot Quantum Dot lasers \cite{Bardella2017}. On the other hand, very few works are dedicated to the study of coherent mode coupling in VCSELs \cite{Prati1994_chaos}. One of the objectives of this paper is to address such a gap in the current literature. While the tool developed in this work can be used to predict the RIN and modulation performance of multimode VCSELs, it also allows for deeper understanding of the physics behind the increase of RIN caused by modal coupling. Further, analytical considerations based on the dynamical model allow to derive the optimal design conditions in terms of mode frequency separation in order to minimize the RIN integrated over the receiver bandwidth and eventually improve the laser bandwidth-power trade-off characteristics. Performing an electromagnetic analysis, which accounts for the detailed structure of a typical mid-area VCSEL, we demonstrate that by altering the oxide aperture ellipticity and axes dimension, we can tailor the modal thresholds and the frequency separation among different modes, thus allowing to identify the most suitable parametric regions for design. Performance analysis with NRZ and PAM4 modulations illustrate the potential application of this laser for bitrates up to 70 Gbit/s. The paper is structured as follows: in Section \ref{sec:exp}, we show the experimental characterization of two different 850 nm multimode VCSELs with elliptical aperture. These two devices, while similar in terms of output power, number of lasing modes, and overall RIN floor, produce very different RIN spectra with, in one case, high RIN peaks in the bandwidth of interest (40\,GHz). In Section \ref{sec:modeling}, we illustrate the numerical tool employed for the accurate simulation of the VCSEL dynamics. In Section \ref{sec:ellipt}, we identify specific frequency separation conditions relevant to optimize the VCSEL performance and conduct a detailed study of the impact of the oxide aperture aspect ratio and overall dimensions on the laser dynamics. 

\section{Experiment}\label{sec:exp}
We consider two different 850 nm few-mode VCSELs with an elliptical oxide aperture, which we identify as A and B. VCSEL A will be used to demonstrate the issue of peaks in the relative intensity noise and exemplify the typical RIN characteristics that we aim to avoid. VCSEL B, exemplifying instead a good device with low RIN in the bandwidth of interest, will be employed to qualitatively address the number and type of transverse modes that characterize the VCSEL emission at different bias currents as well as their thresholds and detunings. Such experimental data will be then used as a starting point for our simulations and detailed analysis of the physics behind the modal dynamics, which is instead difficult to solely convey from an experimental standpoint.\\ 
The VCSELs are contacted through probes for the current supply directly on the chip surface: while this may potentially induce strain effects, the results here presented have been similarly obtained for several chips showing little to no change in the overall modal picture. Light from the VCSEL is coupled to a multimode fiber and measured with a high-resolution optical spectrum analyzer (Yokogawa, AQ6370D). To measure the near-field pattern of the VCSEL transverse modes, the laser far field (obtained through a 20x aspheric objective lens) is directed towards a series of 4 consecutive 750 nm blazed reflective diffraction gratings, searching, for each grating, the first order diffraction beam. Each diffraction grating spatially separates the transverse modes lasing at different wavelengths. The near field at the output of the diffraction gratings (obtained through an infinity-corrected tube lens with focal length of 200 mm) is collected by a polarization resolved camera (Thorlabs, CS505MUP1) with an overall $\approx$ 8 mm x 7 mm imaging area. Such an area allows to easily track the spatial location of transverse modes while changing the bias current. Note that, while a higher number of gratings included in the setup implies a potentially higher resolution, it also implies a reduction of the overall optical power collected at the camera. RIN spectra are acquired with a 40 GHz RIN measurement system (SYCATUS, A0010A-040), comprising a signal analyzer (Keysight Technologies N9030A-544), an optical receiver (SYCATUS) and a digital multimeter (Keysight Technologies 34461A).\\
In Fig. \ref{fig:exp}, for VCSEL A, we show the optical spectra at $I_{bias}=4$ mA (a) and 5 mA (d), as well as the corresponding near-field of the transverse modes measured for the same currents (b,e). The optical spectrum $x$-axis is frequency detuned with respect to the fundamental mode lasing wavelength of about 850 nm at the threshold current. The spatial x-axis of Fig. \ref{fig:exp}(b,e), proportional to frequency due to the effect of gratings, has been linearly scaled to such a frequency detuning in order to facilitate the comparison with the optical spectrum. 
While the experimental setup could potentially allow to measure polarization-resolved near-field patterns,  for the purpose of the present analysis we are uniquely interested in the overall modal spatial topography of the emitted modes. Note that, in the present case, the VCSELs under study emit in one prevalent polarization, effectively suppressing the other polarization and possible associated effects on the laser dynamics and consequently on RIN.\\
Figure \ref{fig:exp}(b,e) display the transverse lasing modes relevant in this current range: their closest analytical approximation is given by the  $\textnormal{TEM}_{00,01,10,02,11}$ Hermite-Gauss modes (which we define as $C_{1,2,3,4,5}$ in the following for the sake of simplicity), emerging in such an order for increasing frequency detunings (referenced to the fundamental mode). Note that in (e) mode $C_5$, while present, fades into the background: for this reason, we display its profile for an increased camera exposure time in the inset, in order to properly display its intensity profile. In Fig. \ref{fig:exp}(c,f) we report the resulting RIN, collected at the same currents. In both figures, the small peak at about 14-15 GHz is associated with relaxation oscillations. In (c) we observe the presence of a peak at about 27\,GHz, while in (f) we observe the presence of three peaks at the frequencies of about 8, 20, and 28 GHz. The fact that such peaks fall within the receiver bandwidth, thus increasing the integrated RIN, is of great concern for applications, when, for example, the VCSEL is directly modulated in PAM2 or PAM4 format with high bitrate data transmission \cite{Tatum2015}. Similar RIN degradation have also been reported in \cite{Quirce2011}. To the best of our knowledge neither the physical origin of these peaks has yet been understood nor methods for avoiding them have been assessed.\\
In Fig. \ref{fig:exp}(g-i), we display the results of the experimental characterization for VCSEL B, in optical spectrum (g), near-field measurement (h), and spectral RIN (i) at a current of 8.5 mA. In (h), we can observe the emergence of modes $C_{1,2,3,4}$, which are in this case the only transverse lasing modes relevant to the VCSEL dynamics in a current range up to 12 mA. For bias currents higher than 12 mA, not relevant for our study, we can observe also modes $C_5$ and $\textnormal{TEM}_{20}$ at higher frequency detunings. 
Self-heating of the VCSEL for increasing currents causes a red shift of the lasing wavelengths \cite{Michalzik2013} that is different for the four transverse modes because of their different transverse spatial profile (see also Fig. \ref{fig:exp}(a,d)). For this reason, we define as reference frequency detuning (with respect to the fundamental mode) the values of detuning measured in the optical spectra at threshold current for each mode. For example, for VCSEL B, these are: $\nu_2=213$ GHz, $\nu_3=292$ GHz, $\nu_4=503$ GHz, respectively for modes $C_{2,3,4}$ with respect to mode $C_1$. Less prominent peaks appearing in Fig. \ref{fig:exp}(g) may be associated either to a small birefringence effect leading to emission of the same mode at the almost suppressed polarization at a slightly different frequency or to other transverse modes (e.g., $\textnormal{TEM}_{11,20}$ below threshold).
From Fig. \ref{fig:exp}(i), we can observe that VCSEL B presents an overall almost flat RIN at both low ($I_{bias}=$3\,mA, in blue) and high ($I_{bias}=$8.5\,mA, in black) currents: this identifies VCSEL B as a good example of a VCSEL where multimode dynamics does not affect the noise performance. In the following, we will apply our model to study the power versus current characteristics and direct current modulation performance of VCSEL B. We will then validate our model against the experimental results of VCSEL B and, based on such a validation and the developed theory, we will explain the worse RIN performance of VCSEL A. Further, we will discuss how a good VCSEL design can be obtained through careful analytical considerations and proper design of the VCSEL oxide aperture.

\section{Modeling}\label{sec:modeling}
We exploit the experimental results obtained for VCSEL B, namely the number and type of lasing transverse modes and their associated detunings, as input parameters in our dynamical simulator to study in depth the impact of modal competition on the laser performance. The simulator is based on the scalar model of \cite{Prati1994_chaos,Prati1994}, modified to include the contribution of carrier diffusion in the transverse plane and accounting for coherent frequency mixing effects and spatial hole burning. As anticipated, we assume in the following that, for the VCSEL under study, emission occurs at one prevalent polarization, as suggested by the experimental characterization, therefore limiting the relevant laser dynamics to one polarization axis. The transverse electric field profile $E(\rho,\phi,t)$ is expanded on an orthonormal set of 4 (real) Hermite-Gauss modes $C_m$ (see Appendix \ref{app:modes}), analytically approximating those measured in Fig. \ref{fig:exp}(h),
\begin{eqnarray}
E_m(t)=\int_0^\infty\int_0^{2\pi} E(\rho,\phi,t)C_m(\rho,\phi) \rho \D\rho \D\phi,
\end{eqnarray}
as a consequence, $E_m(t)$ is the complex temporal component of the electric field for mode $C_m(\rho,\phi)$.
The carrier density in the laser quantum wells $N(\rho,\phi,t)$ is expanded on an orthonormal set of 91 real linear combinations of Gauss-Laguerre modes $B_k$ (see also Appendix \ref{app:modes})
as defined in \cite{Prati1994}
\begin{eqnarray}\label{eq:Nk}
N_k(t)=\int_0^\infty\int_0^{2\pi} N(\rho,\phi,t)B_k(\rho,\phi) \rho \D\rho \D\phi\,,
\end{eqnarray}
where $N_k(t)$ is the $k$-th component of the carrier density distribution in the QW active region.
This allows to focus on the dynamics of the relevant transverse modes that are experimentally observed while also properly simulate the complex carrier transverse profile associated with the bias current spatial variation and spatial hole burning \cite{Debernardi2002}. The choice of the Gauss-Laguerre mode expansion for the carrier density intrinsically allows for a simpler numerical solution of the spatial integral accounting for the carrier diffusion terms (as shown in Appendix \ref{app:diff}). We also highlight that, by definition, modes $C_m$ for $m=1-4$ can be built through linear combinations of modes $B_k$, where, in particular, mode $C_4$ with 3 vertical lobes can be written as $B_{4}(\rho,\phi)/\sqrt{3}-\sqrt{2}B_{5}(\rho,\phi)/\sqrt{3}$, to maintain mode normalization (see Appendix \ref{app:modes}).\\ 
\begin{table}[t!]
\centering
\caption{\bf Parameters for dynamical simulations}
\begin{tabular}{cccc}
\hline
parameter & value & parameter & value \\
\hline
 $\alpha$ & 1 & $\Gamma$ & 0.067\\
$\tau_{p,1}$ & 1.67 ps & $G_N$ & $7.78\times10^{-6}$ cm$^3$/s\\
$\tau_{p,2}$ & 1.2 ps & $\eta_i$ & 0.76\\
$\tau_{p,3}$ & 1.13 ps & V & $7\times 10^{-13}$ cm$^3$\\
$\tau_{p,4}$ & 0.83 ps & $\tau_e$ & 0.92 ns \\
$n_g$ & 3.4 & $\varepsilon$ & $1.71\times 10^{-17}$ cm$^3$\\
 $\bar{D}$ & 30 cm$^2$/s & $N_0$ & $2.37\times 10^{18}$ cm$^{-3}$\\
\hline
\end{tabular}
  \label{tab:param}
\end{table}
The resulting set of differential equations for the electric field and carrier density modal components is the following:
\begin{eqnarray}
&\hspace{-6mm}\frac{\D{E}_m(t)}{\D t}\hspace{-0.5mm}=\hspace{-0.5mm}\left(i\omega_m-\frac{1+i\alpha}{2\tau_{p,m}}\right)\hspace{-0.5mm}E_m(t)+\frac{\Gamma G_N\left(1+i\alpha\right)}{2}f_m(t)+S_{sp}(t) \label{eq:dynE}\\
&\hspace{-6mm}\frac{\D{N}_k(t)}{\D t}\hspace{-0.5mm}=\hspace{-0.5mm}\frac{\eta_i I_k}{eV}-\frac{N_k(t)}{\tau_e}-\frac{n_g^2\epsilon_0G_N}{2\hbar\omega_0}g_k(t)+d_k(t)-4DN_k(t)q_k \label{eq:dynN}
\end{eqnarray}
with
\begin{eqnarray}
    &\hspace{-5mm}f_m(t)=\int_0^\infty\int_0^{2\pi} \frac{E(\rho,\phi,t)}{1+\varepsilon N_p(\rho,\phi,t)}C_m(\rho,\phi)\Delta N(\rho,\phi,t)\rho\D\rho \D\phi\label{eq:fm}\\
    &\hspace{-5mm}g_k(t)=\int_0^\infty\int_0^{2\pi}\frac{|E(\rho,\phi,t)|^2}{1+\varepsilon N_p(\rho,\phi,t)}B_k(\rho,\phi)\Delta N(\rho,\phi,t)\rho\D\rho \D\phi \label{eq:gk}\\
    &\hspace{-5mm}d_k(t)=4D\sum_n N_n(t)\int_0^\infty\int_0^{2\pi} B_k(\rho,\phi)B_n(\rho,\phi)\rho^3\D\rho \D\phi\label{eq:dk}
\end{eqnarray}
where $\Delta N(\rho,\phi,t)=N(\rho,\phi,t)-N_0$, $\rho$ is normalized to the beam waist of the fundamental mode, $\omega_0=2\pi c/\lambda_0$ (with $\lambda_0=850$ nm) is the angular frequency of the fundamental mode $C_1$, and $\omega_m=2\pi\nu_m$ is the angular frequency detuning of  mode $C_m$. The photon lifetime is $\tau_{p,m}$, $\alpha$ is the linewidth enhancement factor, $\Gamma$ is the longitudinal confinement factor, and $G_N=g_N v$ with $g_N$ the differential gain and $v$ the group velocity. The gain compression factor is $\varepsilon$ and $N_0$ is the transparency carrier density. Note that carrier shot noise is here neglected, leaving the spontaneous emission $S_{sp}(t)$ in the electric field equation as the only source of noise, modeled as a Langevin stochastic term \cite{Rimoldi2022}.\\ 
\begin{figure}[t!]
\centering
\includegraphics[width=\columnwidth]{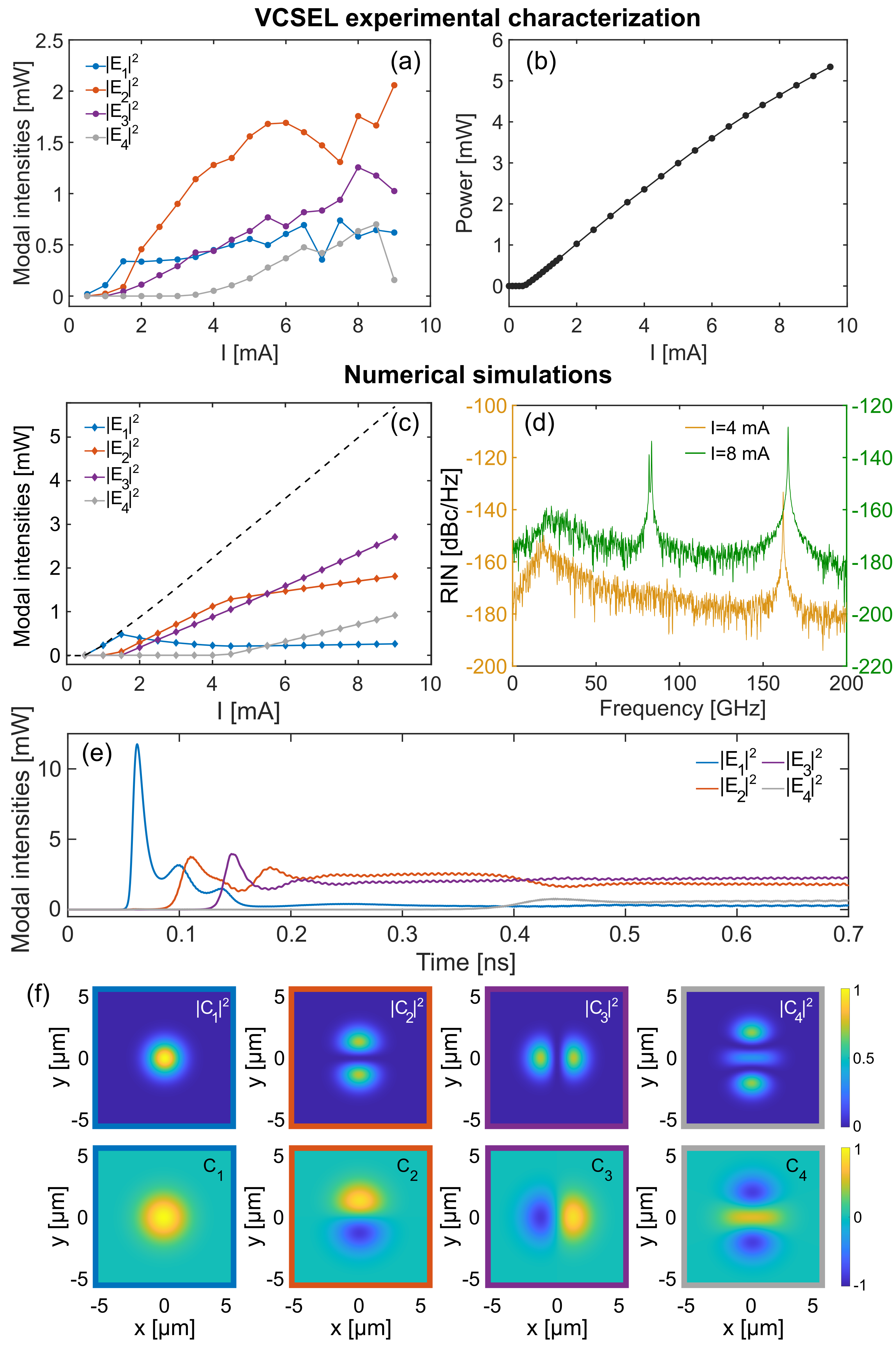}
\caption{(a) Power of the different transverse modes as a function of bias current, extracted from experimentally acquired optical spectra on VCSEL B. (b) Total power as a function of current, measured with a fiber-coupled power-meter on VCSEL B. (c) Numerically simulated modal power as a function of current and resulting total power-current characteristic curve (black dashed line). (d) Simulated spectral RIN at $I=4$ mA (in yellow) and $I=8$ mA (in green). (e) Temporal dynamics of modal intensities for $I=8$ mA. (f) Transverse profiles of the simulated modes in intensity (first row) and corresponding electric field (second row).}
\label{fig:PI_RIN}
\end{figure}
The integral term $f_m(t)$ introduces coupling between mode $C_m$ and the other modes via spatial hole burning of carriers in the active region. Note that, if e.g., the carrier density $N(\rho,\phi,t)$ were constant in the portion of transverse plane where the optical field is different from zero ($N(\rho,\phi,t)=N(t)$) and we assumed the gain compression factor $\epsilon$ to be null, then, due to mode orthogonality, the second term on the RHS of Eq. (\ref{eq:dynE}) would reduce to the standard field amplification term for stimulated emission, written as
\begin{eqnarray*}
&\frac{\Gamma G_N\left(1+i\alpha\right)}{2}f_m(t)\approx\frac{\Gamma G_N\left(1+i\alpha\right)}{2}\left[N(t)-N_0\right]E_m(t)\,.
\end{eqnarray*}
In this case, the time evolution of $E_m(t)$ would be independent from that of the other modes. On the contrary, in our approach, the non-uniform carrier density $N(\rho,\phi,t)$ introduces coupling among the $m$-th mode and the others, with important consequences on the laser dynamics and RIN, as will be analyzed in the following. This approach is innovative with respect to existing models based on rate equations and currently employed to simulate VCSEL performance, since such models neglect coherent coupling among modal fields (present instead in $f_m(t)$ in our case) by considering the sole field intensity (or photon density) instead of the complex electric field \cite{Valle1995,Quirce2011}.\\
In the carrier rate equations (\ref{eq:dynN}), the current injection efficiency is $\eta_i$ and $\tau_e$ is the carrier lifetime, while $V$ is the active region volume. $I_k$ are the modal amplitudes on the $B_k$ basis of the bias pump $I(\rho,\phi)$, for which we assume a super-Gaussian spatial profile \cite{Debernardi2002} and $D$ is the carrier diffusion coefficient $\bar{D}$ multiplied by the square of the beam waist. Further, the integral $g_k(t)$ contains the beating among modes $C_m$ and $d_k(t)$ originates from the carrier diffusion term (see also Appendix \ref{app:diff}). Finally, $N_p(\rho,\phi,t)=n_g^2\epsilon_0|E(\rho,\phi,t)|^2/2\hbar\omega_0$ is the photon density with $n_g$ the group refractive index and $\epsilon_0$ the vacuum permittivity, and $q_k$ is an integer index that identifies families of modes as described in Appendix \ref{app:diff}.\\  
\begin{figure}[t]
\centering
\includegraphics[width=\columnwidth]{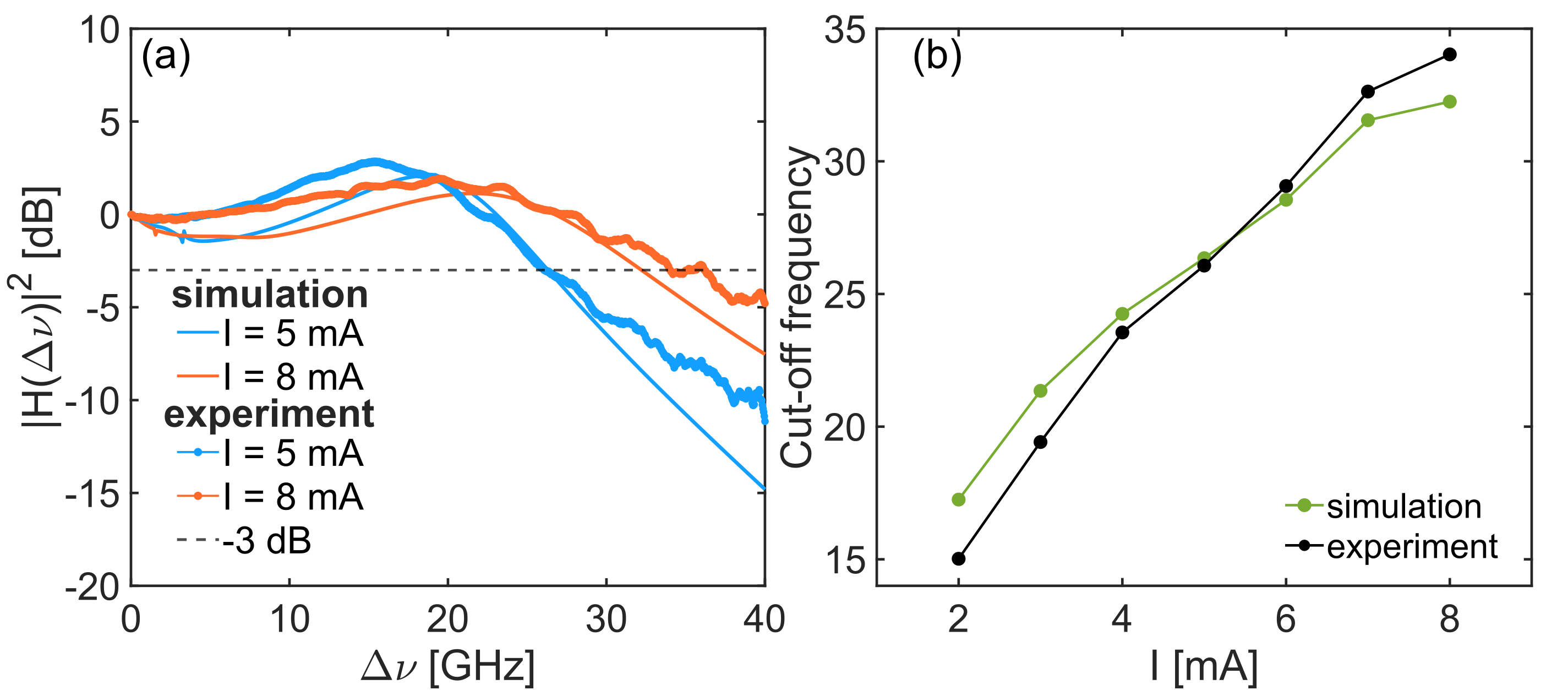}
\caption{(a) Comparison between experimentally obtained (from VCSEL B) and simulated IM responses for bias currents $I=5$ and 8 mA. (b) Plot of the measured (black line) and simulated (green line) laser modulation bandwidth as a function of the bias current.}
\label{fig:IM}
\end{figure}
We report in Table  \ref{tab:param} the relevant parameter values for this study, derived from the fitting of experimental data as well as literature \cite{Coldren,Michalzik2013}. 
In particular, the values of $\tau_{p,m}$ reported in Table \ref{tab:param} have been chosen to fit the experimentally observed laser modal thresholds and modal power-current (PI) curves. This is shown in Fig. \ref{fig:PI_RIN}, where we report (a) the power-current (PI) curve for each mode, extracted from the corresponding peak values in the optical spectra, as well as the total fiber-coupled power as a function of current (b).  
For comparison with the experimental characterization, Fig. \ref{fig:PI_RIN}(c) reports the temporal average of modal intensities versus bias current from simulations with our model, as well as the resulting total output power-current characteristic curve; the matching of (a,b) with (c) shows a good qualitative agreement between simulations and experimental results for increasing current. Figure \ref{fig:PI_RIN}(e) displays the simulated transient of modal intensities when the laser is switched on with a current step of $I=8$ mA, to highlight how modes $C_m$ (in panel (f)) emerge with different time lags and how $C_1$ is partially suppressed by the onset of $C_{2,3}$. Finally, Fig. \ref{fig:PI_RIN}(d) shows the  simulated spectral RIN, at $I=4$ and 8 mA. Focusing on the RIN at 8 mA, the first bump at low frequencies relates to relaxation oscillations, while the peaks at 82, 83, and 165 GHz are due to nonlinear interaction between transverse modes. Note that, in this example, such peaks are beyond a potential bandwidth of the receiver ($\approx 40-50$ GHz) and the bandwidth of the RIN system, hence they cannot be observed in the experimental result of Fig. \ref{fig:exp}(i). 
 The reason behind the formation of these peaks in the spectral RIN, is related to the effect of spatial overlapping of transverse modes: such an overlap is mediated by the non-uniform carrier distribution in the active region, as given by the term $f_m(t)$ in Eq. (\ref{eq:fm}) as will be detailed in the next Section.\\
To demonstrate how the present model can also be employed to simulate the modulation of multimode VCSELs at high bitrates, we report in the following some examples of simulation results for PAM2 and PAM4 modulation.\\
Experimental S-parameter traces are acquired through a microwave network analyzer (Keysight, N5247B PNA-X), where the $S_{21}$ transfer function is properly de-embedded assuming a standard equivalent electrical circuit for the VCSEL structure \cite{Gao2012,Hamad2020}.
In Fig. \ref{fig:IM}(a), we display the comparison between measured and simulated intensity modulation (IM) response, highlighting the good agreement of the -3\,dB bandwidth between experiments and simulations for two values of bias current (5 and 8\,mA). In Fig. \ref{fig:IM}(b) we show the -3\,dB cut-off frequency versus bias current; in particular, at 8\,mA, we observe a bandwidth of 32\,GHz, which is also comparable with other experimental results in literature (see e.g., \cite{Hamed2019}).\\ 
\begin{figure}[t!]
\centering
\includegraphics[width=0.8\columnwidth]{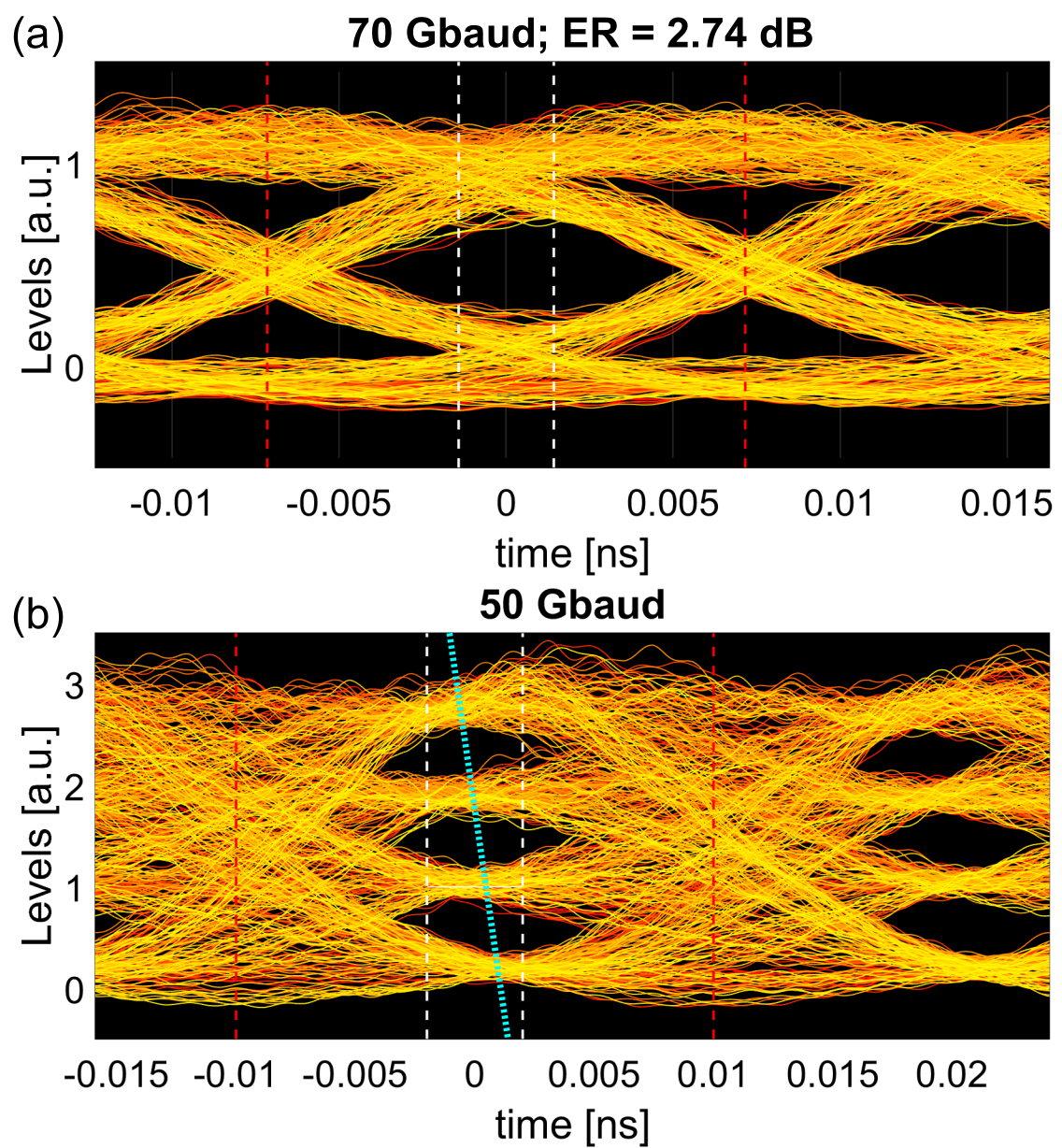}
\caption{Eye diagram for (a) NRZ and (b) PAM4 modulation, both simulated with a PRBS sequence of 1024 bits at an average current $I=8$ mA.}
\label{fig:eye}
\end{figure}
Simulations of PAM2 modulation are carried out with a current injection representing a Non-Return-to Zero (NRZ) Pseudo Random Bit Sequence (PRBS) signal at the average current of 8 mA and peak-to-peak current of 5.3 mA, resulting in an extinction ratio larger than 3 dB. Fig. \ref{fig:eye}(a) displays the resulting eye diagram for a bitrate of 70 Gbit/s and $N_{bit}=2^{10}$, showing an open eye with $\approx$2.74 dB extinction ratio within the 40$\%$-60$\%$ window (marked by the vertical white dashed lines) of a bit period (highlighted by the vertical red dashed lines). This eye diagram is taken at the laser output and does not consider the filtering introduced by the limited bandwidth of the receiver photo-detector. For this reason, we observed that one of the consequences of the $\approx80$\,GHz peaks in the RIN spectrum in Fig. \ref{fig:PI_RIN}(d) consists in increasing the noise on the high and low power levels (more evident on the high level due to the stronger impact of such peaks at higher current, see e.g., Fig. \ref{fig:PI_RIN}(d) at 8 mA). The resulting eye diagram from a PAM4 modulation at a baudrate of 50 Gbaud is displayed in Fig. \ref{fig:eye}(b), for an average current of 8 mA and a current separation between levels of 1.8 mA, showing once again the laser potential capability for high baudrate optical communication. The light blue dotted line in Fig, \ref{fig:eye}(b) highlights the nonlinear eye skews, also seen in experiments of data transmission with very similar VCSELs, which can be compensated with a nonlinear digital predistorter \cite{Minelli2024}.

\section{Discussion and impact of aperture geometry}\label{sec:ellipt}
In this Section, we discuss the physical origin of the high-frequency RIN peaks and, introducing a few reasonable approximations, we derive analytical expressions to calculate their frequency. The final goal is to provide design guidelines for the elliptical oxide aperture, in order to avoid such unwanted frequency components to fall within the receiver bandwidth.
\subsection{Analysis of the high-frequency RIN peaks}
In order to properly discuss the mechanism leading to the RIN peaks at high frequency, we separate  each electric field modal component in Eq. (\ref{eq:dynE}-\ref{eq:dynN}) as $E_m(t)=\tilde{E}_m(t)e^{i\omega_mt}$. In this way, the equation for the complex amplitude $\tilde{E}_m(t)$ becomes
\begin{eqnarray*}
\frac{\D{\tilde{E}}_m(t)}{\D t}=-\frac{1+i\alpha}{2\tau_{p,m}}\tilde{E}_m(t)+\tilde{g}f_m(t)e^{-i\omega_m t}+S_{sp}(t)e^{-i\omega_m t}
\end{eqnarray*}
with $\tilde{g}=\Gamma G_N(1+i\alpha)/2$.
For the sake of simplicity, we assume a spatially constant current profile. We also neglect the carrier diffusion term and set the gain compression factor $\varepsilon$=0. Such simplifications do not affect the calculated frequencies of the RIN peaks but alter slightly their peak intensity with respect to the RIN floor.
Further, in the hypothesis that carriers are fast enough to follow the electric field \cite{Gil2014}, we solve  Eq. (\ref{eq:Nk}) in the adiabatic approximation, which results in
\begin{eqnarray}
&\hspace{-5mm}N(\rho,\phi,t)-N_0\approx\frac{\Lambda}{1+|E(\rho,\phi,t)|^2/E_s^2}\approx \Lambda \left(1-\frac{|E(\rho,\phi,t)|^2}{E_s^2}\right)
\end{eqnarray}
where we set $\Lambda=\tau_e\left(\eta_i I/eV-N_0/\tau_e\right)$ and $1/E_s^2=n_g^2\epsilon_0G_N\tau_e/2\hbar\omega_0$ and assume $|E(\rho,\phi,t)|^2\ll E_s^2$. This procedure reduces the VCSEL dynamics to a set of coupled equations for the electric field modal components $\tilde{E}_m$ where the terms $f_m(t)e^{-i\omega_m t}$, accounting for modal competition, are now approximated as
\begin{eqnarray}\label{eq:fmfirstapprox}
\hspace{-10mm}f_m(t)e^{-i\omega_m t}\approx&\Lambda\int_0^\infty \int_0^{2\pi}  E(\rho,\phi,t)C_m(\rho,\phi)\cdot\nonumber\\ 
&\cdot\left(1-\frac{|E(\rho,\phi,t)|^2}{E_s^2}\right) e^{-i\omega_m t} \rho d\rho d\phi
\end{eqnarray}
Expanding the electric field on its modal components and, in particular,
\begin{eqnarray*}
|E(\rho,\phi,t)|^2=\sum_{l,r} \tilde{E}_l(t)\tilde{E}_r(t)^* e^{i\omega_{lr}t}C_l(\rho,\phi)C_r(\rho,\phi)
\end{eqnarray*}
with $\omega_{lr}=\omega_l-\omega_r$, we obtain
\begin{eqnarray} \label{eq:fmapprox}
&\hspace{-48mm}f_m(t)e^{-i\omega_mt}=\Lambda\left[\tilde{E}_m(t)\right.\nonumber\\
&\hspace{10mm}\left.-\sum_{n,l,r} \gamma_{mnrl}\frac{\tilde{E}_n(t)\tilde{E}_l(t)\tilde{E}_r^*(t)}{E_s^2}e^{i(\omega_{nm}+\omega_{lr})t}\right]
\end{eqnarray}
where the beating frequencies are given by $\omega_{nm}+\omega_{lr}$, and
\begin{eqnarray*}
\gamma_{mnrl}=\int_0^\infty \int_0^{2\pi}C_m(\rho,\phi)C_n(\rho,\phi)C_r(\rho,\phi)C_l(\rho,\phi)\rho \D\rho \D\phi&
\end{eqnarray*}
We highlight that the terms $\gamma_{mnrl}$ account only for the overlapped spatial profiles of the interacting modes and, similarly to \cite{Lenstra2014} in the case of edge-emitting lasers, they can be calculated a priori, given the expression of transverse modal profiles $C_m(\rho,\phi)$ (reported in Appendix \ref{app:modes}). It is evident that the $\gamma_{mnrl}$ coefficients always give the same value for all possible permutations of the same considered 4 modes ($m,n,r,l$). On the other hand, the corresponding beating frequencies $\omega_{nm}+\omega_{lr}$ depend on the considered permutation (see Appendix \ref{app:coeff}). The number of $\gamma_{mnrl}$ coefficients different from each other for the entire set of equations is given by the binomial coefficient $\binom{M+4-1}{4}$ with $M$ being the number of considered modes ($M=4$ in our case). We can further isolate in Eq. (\ref{eq:fmapprox}) the cases where $\omega_{nm}+\omega_{lr}=0$. The resulting equation for the dynamics of modal components (neglecting the stochastic spontaneous emission contribution) is
\begin{eqnarray}
&\hspace{-20mm}\frac{\D{\tilde{E}}_m}{\D t}=-\frac{1+i\alpha}{2\tau_{p,m}}\tilde{E}_m+\tilde{g}\Lambda \left[1-\gamma_{mmmm}\frac{|\tilde{E}_m|^2}{E_s^2}\right.\label{eq:simplEm}\\
&\hspace{-22mm}\left.-2\sum_{n\neq m}\gamma_{mmnn}\frac{|\tilde{E}_n|^2}{E_s^2}\right]\tilde{E}_m\label{eq:othersat}\\
&\hspace{-5mm}-\tilde{g}\Lambda\sum_{n,r,l*}\gamma_{mnrl}\frac{\tilde{E}_n\tilde{E}_l\tilde{E}_r^*}{E_s^2}e^{i(\omega_{nm}+\omega_{lr})t}\,,\label{eq:beatterm}
\end{eqnarray}
where the sum $\sum_{n,r,l*}$ is extended to all indices combinations excluded those already extracted in (\ref{eq:simplEm}-\ref{eq:othersat}) for $n-m+l-r=0$. We observe that the RHS of (\ref{eq:simplEm}) represents the standard contribution of the cavity loss, $\alpha$ factor, stimulated emission, and self-saturation; term (\ref{eq:othersat}) is the cross-saturation that compresses the power of the $m$-th mode (via spatial hole burning) when another $n$-th mode turns on (see e.g., Fig. \ref{fig:PI_RIN}(a) and (c)). The last term (\ref{eq:beatterm}) accounts for the four-wave mixing that causes oscillations in the electric field of the $m$-th mode at beating frequency $\omega_{n}-\omega_{m}+\omega_{l}-\omega_{r}$.
We further highlight that not all the beating frequency combinations in the term (\ref{eq:beatterm}) will effectively appear in the temporal dynamics of $\tilde{E}_m$, but only those for which $\gamma_{mnrl}$ is non-null. In fact, it is important to point out that, if the integration domain is symmetric with respect to the origin for at least one integration variable (as is the case for modes $C_m$ with respect to $x$ and $y$) and if the integrand is odd with respect to such a variable, then $\gamma_{mnrl}$ is zero (see Appendix \ref{app:coeff}). This property results very useful in simplifying the overall form of the equations.\\ 
The total output power of the VCSEL measured at the photo-detector is
\begin{eqnarray*}
&P_{tot}(t)=\int_{0}^\infty\int_{0}^{2\pi}\left|\sum_m \tilde{E}_m(t)e^{i\omega_mt}C_m(\rho,\phi)\right|^2\hspace{-2mm}\rho\D\rho \D\phi\\
&\hspace{-28mm}=\sum_m\left|\tilde{E}_m(t)\right|^2\,,
\end{eqnarray*}
with temporal evolution governed by the differential equation $\D P_{tot}(t)/\D t=\sum_m \D|\tilde{E}_m|^2/\D t$.  Hence, the beating frequencies observed in the dynamics of $|\tilde{E}_m|^2$ are bound to appear also in the power dynamics and in the RIN. All beating terms in expression (\ref{eq:beatterm}) and the corresponding beating frequencies for the case of 4 transverse modes are reported in Tables \ref{tab:gamma} and \ref{tab:beatings} of Appendix \ref{app:coeff}.\\ 
\begin{figure}[t]
\centering
\includegraphics[width=0.75\columnwidth]{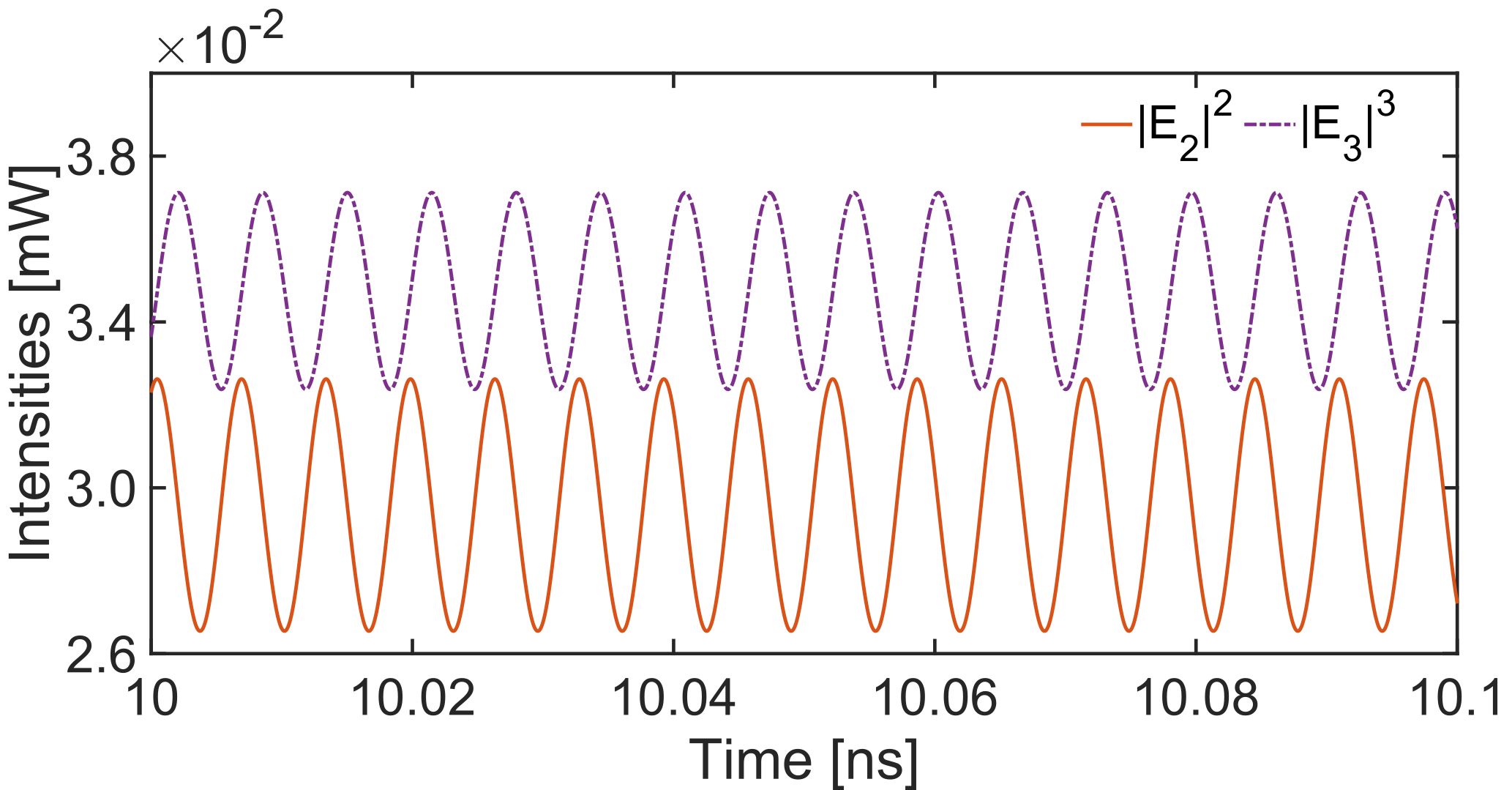}
\caption{Temporal evolution of the intensity of modal components $E_{2,3}$ near the laser threshold, simulated under the approximations illustrated in Section \ref{sec:ellipt}. The oscillation frequency of both variables ($\approx$155 GHz) approximates the second harmonic of their modal frequency separation $\nu_3-\nu_2=79$ GHz.}
\label{fig:2modes}
\end{figure}
For the device simulated in Fig. \ref{fig:PI_RIN}, the beating terms with non-null $\gamma_{mnrl}$ coefficients falling within a 200 GHz window are: (i) $\pm 2(\omega_3-\omega_2)$, with coefficients, e.g., $\gamma_{2323}$, caused by the coupling between $C_2$ and $C_3$, (ii) $\pm(\omega_4-2\omega_{2})$, with coefficients, e.g., $\gamma_{2124}$, caused by the coupling of $C_4$, $C_2$, and $C_1$, and (iii) $\pm(\omega_4-2\omega_{3})$, with coefficients, e.g., $\gamma_{3134}$, caused by the coupling of $C_4$, $C_3$, and $C_1$. This explains the peaks at high frequencies observed in the simulated spectral RIN in Fig. \ref{fig:PI_RIN}(d): the peak at 165 GHz is close to $2|\nu_3-\nu_2|$, while the two peaks at 82 and 83 GHz are given by $|\nu_4-2\nu_2|$ and $|\nu_4-2\nu_3|$, respectively. We highlight in particular that the beatings at 82 and 83 GHz appear only when mode $C_4$ is above threshold (e.g., for $I=8$ mA). Note that the coupling of $C_2$ and $C_3$ gives only a non-null second harmonic because the first harmonic at $|\nu_2-\nu_3|$ has coefficients, e.g., $\gamma_{2322}=0$, due to the odd symmetry of the product of the coupled transverse mode profiles. These considerations lead to the conclusion that a VCSEL design (with elliptical oxide aperture and four lasing modes as in Fig. \ref{fig:exp}(h)) that minimizes the RIN must guarantee frequency detunings among transverse modes such that these beatings remain outside the receiver bandwidth BWR, that is:
\begin{eqnarray}\label{eq:beatingcond}
2|\nu_3-\nu_2|>\textrm{BWR}  \quad , \quad |\nu_4-2\nu_{2,3}|>\textrm{BWR}
\end{eqnarray}
To further explore this concept, we analyze a simplified case where only modes $C_2$ and $C_3$ are lasing. Here, the dynamics of $|E_2|^2$ follows the equation
\begin{eqnarray*}
&\hspace{-27mm}\frac{\D|\tilde{E}_2(t)|^2}{\D t}=\tilde{E}_2(t)\frac{\D\tilde{E}_2^*(t)}{\D t}+\tilde{E}_2^*(t)\frac{\D\tilde{E}_2(t)}{\D t}\,\Leftrightarrow\\
&\hspace{-35mm}\frac{\D|\tilde{E}_2(t)|^2}{\D t}=(-\frac{1}{\tau_{p2}}+\Gamma G_N\Lambda)|\tilde{E}_2(t)|^2
\\
&\hspace{5mm}-\frac{\Lambda}{E_s^2}\Gamma G_N\left[\gamma_{2222}|\tilde{E}_2(t)|^4+2\gamma_{2233}|\tilde{E}_2(t)|^2|\tilde{E}_3(t)|^2\right]\\
&\hspace{7mm}-\frac{\Lambda}{E_s^2}\Gamma G_N\gamma_{2323}|\tilde{E}_2^*(t)\tilde{E}_3(t)|^2\left\{\cos[\theta(t)]+\alpha\sin[\theta(t)]\right\}
\end{eqnarray*}
with $\theta(t)=2[\phi_3(t)-\phi_2(t)]+2(\omega_3-\omega_2)t$, which would lead to a temporal oscillation of $|\tilde{E}_2|^2$. $\phi_2(t)$ and $\phi_3(t)$ are the additional phase terms for the complex electric field modal components, defined as $\tilde{E}_2(t)=|\tilde{E}_2(t)|e^{i\phi_2(t)}$ and $\tilde{E}_3(t)=|\tilde{E}_3(t)|e^{i\phi_3(t)}$. A similar equation can also be found for the dynamics of $|\tilde{E}_3(t)|^2$. By numerically simulating the set of two coupled equations in Eq. (\ref{eq:simplEm}-\ref{eq:beatterm}) for the complex amplitudes of modes $C_{2}$ and $C_3$, we observe what is reported in Fig. \ref{fig:2modes}, where the only beating frequency displayed is indeed approximately at the second harmonic of the frequency separation between the two transverse modes \cite{Prati1994}. The residual difference in phase given by $2[\phi_3(t)-\phi_2(t)]$ gives rise to an instantaneous frequency that does not exceed $\approx 2$ GHz, after the initial transient dynamics and can thus be neglected with respect to $2(\omega_3-\omega_2)t$.\\
\begin{figure}[t]
\centering
\includegraphics[width=\columnwidth]{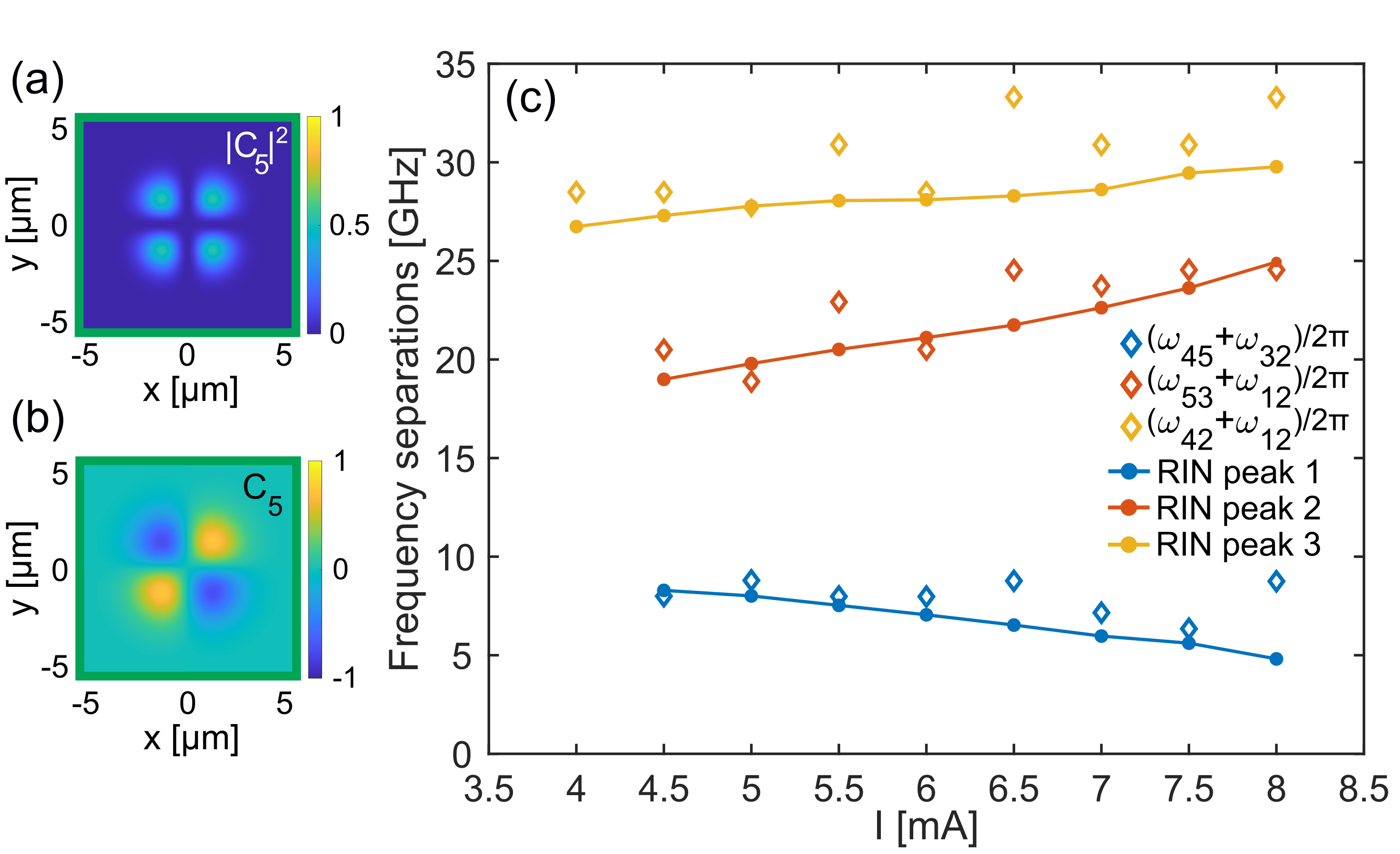}
\caption{Transverse profile of mode $C_5$ in intensity (a) and electric field (b). (c) Trend of the experimental RIN peaks (solid line with point markers) and corresponding detuning condition (diamond markers) calculated from the experimental optical spectra for different currents.}
\label{fig:est_freq_sep}
\end{figure}
We now return to the experimental results reported in Fig. \ref{fig:exp} for VCSEL A. Applying the same methodology, we are now able to address the peaks observed in the spectral RIN in Fig. \ref{fig:exp}(c,f). The frequency detunings of modes $C_{2,3,4}$ with respect $C_1$ differ from those of VCSEL B: in particular, we have approximately $\nu_2=189.6$\,GHz, $\nu_3=257.4$\,GHz, $\nu_4=407.6$\,GHz in Fig. \ref{fig:exp}(a) at $I_{bias}=4$\,mA. The single peak at 27 GHz in Fig. \ref{fig:exp}(c) is then due to the beating $\nu_4-2\nu_2=28.4$ GHz highlighted in Eq. (\ref{eq:beatingcond}), while the other beating frequencies in Eq. (\ref{eq:beatingcond}) are beyond the receiver bandwidth at 40\,GHz.\\ 
On the other hand, in Fig. \ref{fig:exp}(d-f), we notice that, differently from VCSEL B, mode $C_5$ is already above threshold at $I_{bias}=5$\,mA, due to the different internal design of VCSEL A. The modal profile of $C_5$ is reported in Fig. \ref{fig:est_freq_sep}(a,b) in intensity and electric field, respectively. Here, our theoretical framework can still be applied in the case of 5 relevant transverse modes. From Fig. \ref{fig:exp}(d), we have $\nu_2=196.7$\,GHz, $\nu_3=266.8$\,GHz, $\nu_4=422.6$\,GHz, $\nu_5=484$\,GHz at $I_{bias}=5$\,mA. Apart from the beating at $\nu_4-2\nu_2=29.2$ GHz resulting in the peak at the highest frequency in Fig. \ref{fig:exp}(f), our theory predicts only two other beating frequencies falling within the BWR that match the other two peaks reported n Fig. \ref{fig:exp}(f) (see Tables \ref{tab:M5gamma}, \ref{tab:M5beatings} in Appendix \ref{app:coeff} for more details). The first peak at low frequencies is due to the beating at $\nu_4-\nu_5+\nu_3-\nu_2=8.7$\,GHz, with associated coefficient $\gamma_{5423}$. Note that $\gamma_{5423}$ is not null because both products of $C_4$ and $C_5$, on one side, and of $C2$ and $C_3$, on the other side, give a function that is odd in both $x$ and $y$, resulting in an even function in both $x$ and $y$ for the product of the four modes and therefore a non-null integral. The second peak is instead due to the beating frequency $\nu_5-\nu_3+\nu_1-\nu_2=20.5$ GHz, i.e., $\nu_5-\nu_3-\nu_2$ in our reference with $\nu_1=0$, with associated coefficient $\gamma_{3521}$, which is also a non-null integral for similar reasons. In Fig. \ref{fig:est_freq_sep}(c) we report the measured experimental RIN peaks (solid line with point markers) and compare them with the detuning conditions calculated from the experimental optical spectra (diamond markers) for a current range between 4 and 8 mA. Here, yellow diamond markers refer to the beating at $\nu_4-2\nu_2$, orange diamond markers refer to the beating at $\nu_5-\nu_3-\nu_2$, and blue diamond markers refer to the beating at $\nu_4-\nu_5+\nu_3-\nu_2$. Apart from the inherent minor precision of the OSA measurement, we can observe a very good agreement between the predicted peaks and those observed in the experimental RIN.\\
We finally highlight that most of the beating frequencies in Eq. (\ref{eq:simplEm}-\ref{eq:beatterm}) emerge due to the correct inclusion of coherent frequency mixing in the present model: if, instead of $|E(\rho,\phi,t)|^2$ in Eq. (\ref{eq:fmfirstapprox}), we were to consider only the incoherent sum $\sum_{m=1}^4|E_mC_m|^2$, as often performed in literature on multimode dynamics (see e.g., in \cite{Law1997}), this would result only either in a beating frequency at $\nu_4$ or in no predicted beating \cite{Valle1995} (see Appendix \ref{app:incorrect_mixing}). We conclude that, if we neglect the coherent modal beating in the carrier density, we cannot replicate the RIN peaks as seen in the experiment of Fig. \ref{fig:exp}(c,f) or in the numerical simulations of Fig. \ref{fig:PI_RIN}(d).
\subsection{Influence of aperture geometry on beating frequencies}
\begin{figure*}[t]
\centering
\includegraphics[width=0.7\textwidth]{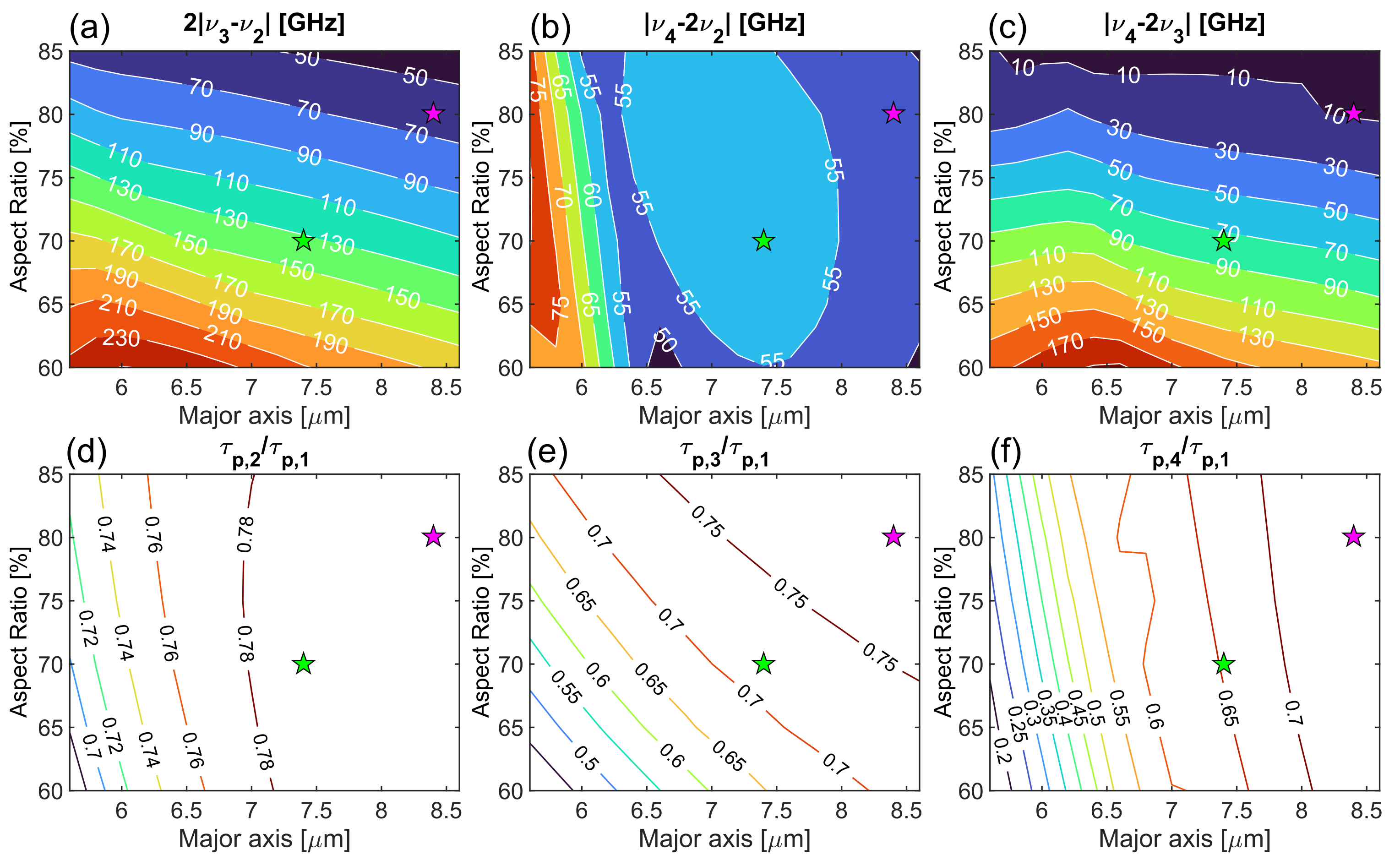}
\caption{(a-c) Maps of detuning conditions as identified by Eq. (\ref{eq:beatingcond}) as a function of the oxide aperture aspect ratio ($y$-axis) and ellipse major axis ($x$-axis). (a) Second harmonic of the beating frequency between modes $C_2$ and $C_3$. (b) Frequency separation between the frequency detuning $\nu_4$ and the second harmonic of the detuning $\nu_2$. (c) Frequency separation between $\nu_4$ and the second harmonic of the detuning $\nu_3$. (d-f) Ratios of modal photon lifetimes for modes $C_{2-4}$ respectively, over the photon lifetime of the fundamental mode.}
\label{fig:detcond}
\end{figure*}
In the final part of this paper, we address the impact of the oxide aperture size, as well as the aperture ellipse aspect ratio on the laser dynamics. As illustrated in the previous discussion, more than the frequency separation between different transverse modes, it is important to study the behavior of specific beating frequencies generated through complex modal competition, namely given by Eq. (\ref{eq:beatingcond}) for an elliptically-shaped oxide aperture with four relevant modes. These are indeed the frequency separation that may result in a peak in the spectral RIN, with a chance of degrading the VCSEL performance. VCSELs are complex structures, composed of hundreds of layers, which can be planar or transversally modulated, as in the case of the oxide aperture. Our reference VCSEL features an almost 100$\%$ reflectivity bottom DBR of 41 graded interface pairs, an out-coupling top mirror of 22 pairs, and an oxide included in the first pair above the l-cavity, which embeds 3 GaAs QWs.
In the following, we investigate the VCSEL through our in-house vectorial and three-dimensional VCSEL solver VELMS (Vcsel ELectroMagnetic Suite) \cite{Bava2001, Debernardi2003}, where the oxide aperture and gain region are elliptical, and to optimize the optical emission spectrum in order to fulfill the conditions set by Eq. (\ref{eq:beatingcond}). VELMS expands the VCSEL transverse modes in terms of the complete vectorial modal basis of a reference medium in cylindrical coordinates. Applying the reciprocity theorem of Maxwell equations, the method can propagate the expansion coefficients within each layer of any transverse size and shape. VCSEL modes are then determined by requiring that their expansion coefficients replicate themselves after a full resonator roundtrip (Barkhausen criterion). This defines mode wavelengths and modal losses, or photon lifetimes, by finding the QW gain that compensates for material, diffraction, and radiation losses.  In the present case, VELMS outputs of interest are the emission wavelengths, as well as their corresponding photon lifetimes and field topographies, both at the level of the active region (for potential use as input parameters for our dynamical model), and at the output facet, allowing comparison with the experimental near-field results in Fig. \ref{fig:exp}(h). 
Using the modal wavelengths computed by VELMS, in Fig. \ref{fig:detcond}(a-c), we report the maps of the relevant beating frequencies for different elliptical oxide apertures. These are parametrized by the ellipse major axis $a$ ($x$-axis) and the relative size $b/a$ (aspect ratio, expressed in percentage on the $y$-axis) of the minor axis $b$. An aspect ratio (AR) of 100$\%$ (not displayed) describes a perfectly circular oxide aperture, characterized by frequency degeneracy for modes with the same radial dependence, such as $C_2$ and $C_3$. 
While Fig. \ref{fig:detcond}(a-c) show the behavior of the harmonic combinations given by Eq. (\ref{eq:beatingcond}), Fig. \ref{fig:detcond}(d-f) show the corresponding $\tau_{p,m}/\tau_1$ ($m=2-4$), where $\tau_{p,m}$ are the modal photon lifetimes. A low ratio indicates that the considered mode is unlikely to turn on above threshold after the lasing of $C_1$.\\
Maps in Fig. \ref{fig:detcond}(a-c) can be used to identify good designs that satisfy the constraints of Eq. (\ref{eq:beatingcond}). Assuming e.g., BWR\,$=50$ GHz, a good design is highlighted by the green star in Fig. \ref{fig:detcond}(a-c). Smaller ellipse major axes (e.g., 6 $\mu$m or less) equally satisfy Eq. (\ref{eq:beatingcond}) but at the expense of a much lower emitted power for the same bias current \cite{Moser2015,Michalzik2013}.\\
The violet star marker (at AR of 80$\%$ and major axis 8.4 $\mu$m) identifies a region of high power but potentially displaying RIN peaks within BWR (see Fig. \ref{fig:detcond}(b,c)). In particular, in this case, at least one of the predicted beating frequencies, i.e., the one at about 10 GHz for condition (c), is expected to lie within the receiver bandwidth.\\ 
Next, we focus on the region highlighted by the green star at AR of 70$\%$ and ellipse major axis 7.4 $\mu$m. This overall area (for ARs between $\approx$ 65 and 75$\%$ and major axes between $\approx$ 6.8 and 7.8 $\mu$m) is characterized by a good trade-off between the high-power requirement and the maximization of beating frequencies. In particular, we observe that, while all four modes can lase at e.g., 5 mA and above (see (d-f)), the maps in Fig. \ref{fig:detcond}(a-c) indicates all peaks in the spectral RIN to be beyond the receiver bandwidth, similarly to the experimental and numerical case illustrated in Fig. \ref{fig:exp}(i) and Fig. \ref{fig:PI_RIN}(d), respectively.\\
\begin{figure}[t]
\centering
\includegraphics[width=\columnwidth]{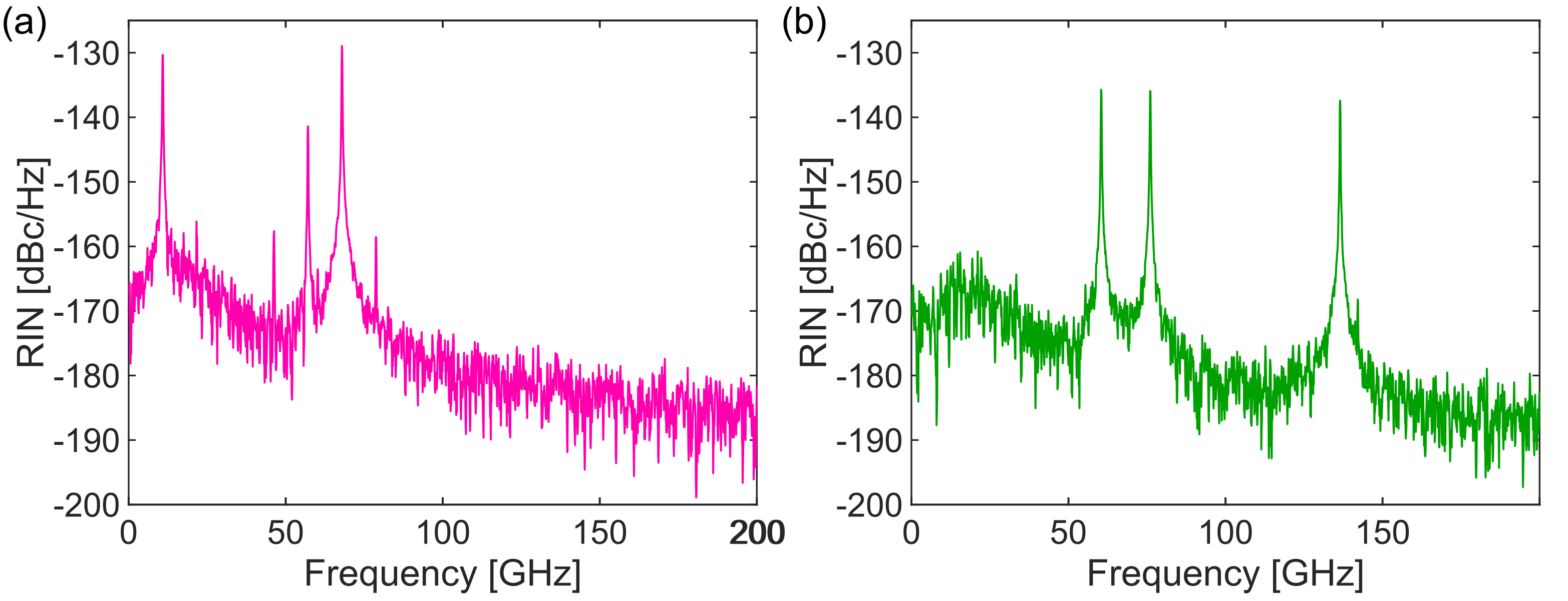}
\caption{(a,b) Spectral RIN obtained through dynamical simulations for parameters matching the star markers in Fig. \ref{fig:detcond} at $I=5$ mA.}
\label{fig:RINstar}
\end{figure}
In order to validate these considerations, we performed dynamical simulations, based on Eq. (\ref{eq:dynE}-\ref{eq:dk}), for the parametric regions identified by the star markers at a fixed bias current $I=5$ mA. We report the resulting spectral RIN in these two cases in Fig. \ref{fig:RINstar}(a,b), corresponding to ARs and ellipse major axes of (a) 80$\%$, 8.4 $\mu$m and (b) 70$\%$, 7.4 $\mu$m. In the case of Fig. \ref{fig:RINstar}(a) the electromagnetic solver predicts beating frequencies at about 61 (Fig. \ref{fig:detcond}(a)), 52 (Fig. \ref{fig:detcond}(b)), and 9 (Fig. \ref{fig:detcond}(c)) GHz, which are all present when running the numerical simulation at 5 mA. The slight difference in the peak frequency values in Fig. \ref{fig:RINstar}(a) (at 68, 57, 11 GHz) is due to the fact that the electromagnetic solver predicts the modal detuning frequencies at threshold for each mode. Here, we expect the peak at about 11 GHz to be the most limiting to the laser performance since, as in the case of Fig. \ref{fig:exp}(c,f), it would lie within receiver bandwidths and would lead to an overall increase of the integrated RIN (RIN\,$=-150$\,dBc integrated over a bandwidth of 50 GHz). Finally, Fig. \ref{fig:RINstar}(b) identifies the most promising high-power case: here, all beating frequencies (at 60, 76, and 136 GHz) lie beyond the receiver bandwidth at about 50 GHz, which results in a low integrated RIN (RIN\,$=-167$\,dBc for a 50 GHz bandwidth).
\section{Conclusions}
We have studied the properties of a multimode 850 nm VCSEL with elliptical aperture. Based on experimental results for the modal profiles, thresholds, and frequency detunings, we investigated the laser dynamics and RIN behavior, employing a multimode and spatially-resolved dynamical model that accounts for coherent effects and spatial hole burning.\\
Experimental results show a fast modulation response, suitable for high-speed datacom applications, but limited by RIN issues, which we have been able to relate, for the first time, to transverse mode coupling.
Aiming at datacom applications, a detailed analysis of the laser dynamics has allowed to identify the main conditions on the modal detuning frequencies to optimize the relative intensity noise, which have been verified against experimental results and can be controlled by addressing the oxide aperture aspect ratio and dimensions. Investigating these quantities through an advanced electromagnetic solver, we have been able to identify the best parametric region for the design of these structures, allowing for an optimized bandwidth-power trade-off. 

\section*{Acknowledgments}
The authors acknowledge Dr. Fabrizio Forghieri (Cisco Photonics, Vimercate, Italy) for coordinating the project under a CISCO Sponsored Research Agreement. This work was also partially supported by the European Union under the Italian National Recovery and Resilience Plan (PNRR) of NextGenerationEU, partnership on "Telecommunications of the Future" (PE00000001 - program "RESTART"). CR acknowledges funding from research contract no. [32-I-13427-1] (DM 1062/2021) funded within the Programma Operativo Nazionale (PON) Ricerca e Innovazione of the Italian Ministry of University and Research.

{\appendices
\section{Analytical expression of transverse modal profiles}\label{app:modes}
Following \cite{Prati1994}, we define modes $B_k(\rho,\phi)$ as a real orthonormal basis given by linear combinations of orthonormal Gauss-Laguerre modes $A_{p,l}(\rho,\phi)$. We start by defining
\begin{eqnarray*}
A_{p,l}(\rho,\phi)=\sqrt{\frac{2}{\pi}}(2\rho^2)^{|l|/2}\left[\frac{p!}{(p+|l|)!}\right]^{1/2}\hspace{-1mm}L_p^{|l|}(2\rho^2)e^{-\rho^2}e^{il\phi}
\end{eqnarray*}
where $p$ is the radial index, $l$ is the angular index and $L_p^{|l|}(2\rho^2)$ are the Laguerre polynomials of argument $2\rho^2$. Modes $B_{p,l,n}(\rho,\phi)$ (or $B_k(\rho,\phi)$ in short notation) are then defined as \cite{Prati1994} 
\begin{align*}
B_{p,0}(\rho,\phi)&=A_{p,0}(\rho,\phi)\\
B_{p,l,1}(\rho,\phi)&=\frac{1}{\sqrt{2}}\left[A_{p,l}(\rho,\phi)+A_{p,-l}(\rho,\phi)\right]\\
&=\frac{2}{\sqrt{\pi}}(2\rho^2)^{l/2}\left[\frac{p!}{(p+l)!}\right]^{1/2}\hspace{-1mm}L_p^l(2\rho^2)e^{-\rho^2}\cos{(l\phi)}\nonumber\\
B_{p,l,2}(\rho,\phi)&=\frac{1}{\sqrt{2}i}\left[A_{p,l}(\rho,\phi)-A_{p,-l}(\rho,\phi)\right]\\
&=\frac{2}{\sqrt{\pi}}(2\rho^2)^{l/2}\left[\frac{p!}{(p+l)!}\right]^{1/2}\hspace{-1mm}L_p^l(2\rho^2)e^{-\rho^2}\sin{(l\phi)}\nonumber
\end{align*}
We introduce \cite{Prati1994} in short notation $B_1=B_{0,0}$, $B_2=B_{0,1,1}$, $B_3=B_{0,1,2}$, $B_4=B_{1,0}$, $B_5=B_{0,2,1}$, $B_6=B_{0,2,2}$.
Finally, we define the first six modes of the orthonormal basis of Hermite-Gauss modes $C_m(\rho,\phi)$ as linear combinations of $B_{p,l,n}$. In particular
\begin{align*}
C_1(\rho,\phi)&=B_1(\rho,\phi)=\sqrt{\frac{2}{\pi}}e^{-\rho^2}\\
C_2(\rho,\phi)&=B_2(\rho,\phi)=\frac{1}{\sqrt{2}}\left[A_{0,1}(\rho,\phi)+A_{0,-1}(\rho,\phi)\right]\nonumber\\
&=\sqrt{\frac{2}{\pi}}2\rho\cos{\phi}\,e^{-\rho^2}\\
C_3(\rho,\phi)&=B_3(\rho,\phi)=\frac{1}{\sqrt{2}i}\left[A_{0,1}(\rho,\phi)-A_{0,-1}(\rho,\phi)\right]\nonumber\\
&=\sqrt{\frac{2}{\pi}}2\rho\sin{\phi}\,e^{-\rho^2}\\
C_4(\rho,\phi)&=\frac{1}{\sqrt{3}}B_4(\rho,\phi)-\sqrt{\frac{2}{3}}B_5\\
&=\frac{1}{\sqrt{3}}\left[A_{1,0}(\rho,\phi)-A_{0,2}(\rho,\phi)-A_{0,-2}(\rho,\phi)\right]\nonumber\\
C_5(\rho,\phi)&=B_6(\rho,\phi)=\sqrt{\frac{2}{\pi}}2\rho^2\sin{(2\phi)}e^{-\rho^2}\\
C_6(\rho,\phi)&=\frac{1}{\sqrt{3}}B_4(\rho,\phi)+\sqrt{\frac{2}{3}}B_5\\
&=\frac{1}{\sqrt{3}}\left[A_{1,0}(\rho,\phi)+A_{0,2}(\rho,\phi)+A_{0,-2}(\rho,\phi)\right]\nonumber
\end{align*}
In a more common notation, $C_1=\textnormal{TEM}_{00}$, $C_2=\textnormal{TEM}_{01}$, $C_3=\textnormal{TEM}_{10}$, $C_4=\textnormal{TEM}_{02}$, $C_5=\textnormal{TEM}_{11}$, and $C_6=\textnormal{TEM}_{20}$.

\section{Derivation of diffusion term}\label{app:diff}
In Eq.\,(\ref{eq:dynN}) describing the dynamics of the carrier density modal component $k$, the term
\begin{align}\label{eq:dkapp}
&d_k(t)-4DN_k(t)q_k=\\
&=4D\sum_n N_n(t)\int_0^\infty\int_0^{2\pi} B_k B_n\rho^3\D\rho \D\phi-4DN_k(t)q_k\nonumber
\end{align}
 accounts for the carrier diffusion: in the following we illustrate its derivation. In particular, following a known property of Gauss-Laguerre modes \cite{Prati1994}, if we express the transverse Laplacian operator $\nabla_{\bot}^2$ in polar coordinates and use the expression of Gauss Laguerre modes $A_{p,l}$, defined in the previous Section, we obtain:
\begin{eqnarray}
\nabla_{\bot}^2 A_{p,l}(\rho,\phi)=4\left[\rho^2-(2p+|m|+1)\right]A_{p,l}(\rho,\phi)
\end{eqnarray}
The same relation holds for modes $B_{k}=B_{p,l,n}=B_{p,l,n}(\rho,\phi)$, since they are linear combinations of modes $A_{p,l}=A_{p,l}(\rho,\phi)$:
\begin{align}
\left(\frac{1}{4}\nabla_{\bot}^2-\rho^2\right) B_{p,l,0}&=\left(\frac{1}{4}\nabla_{\bot}^2-\rho^2\right) A_{p,0}\nonumber\\
&=-(2p+|m|+1)B_{p,l,0}\label{eq:Brel1}\\
\left(\frac{1}{4}\nabla_{\bot}^2-\rho^2\right) B_{p,l,1}&=\left(\frac{1}{4}\nabla_{\bot}^2-\rho^2\right) \frac{1}{\sqrt{2}}\left[A_{p,l}-A_{p,-l}\right]\nonumber\\
&=-(2p+|m|+1)B_{p,l,1}\label{eq:Brel2}\\
\left(\frac{1}{4}\nabla_{\bot}^2-\rho^2\right) B_{p,l,2}&=\left(\frac{1}{4}\nabla_{\bot}^2-\rho^2\right) \frac{1}{\sqrt{2}i}\left[A_{p,l}-A_{p,-l}\right]\nonumber\\
&=-(2p+|m|+1)B_{p,l,2}\label{eq:Brel3}
\end{align}
The diffusion term in Eq. (\ref{eq:dkapp}) for the modal components $N_k$, results from the expansion of the carrier density $N$ on the $B_k$ orthonormal basis and the use of Eq. (\ref{eq:Brel1}-\ref{eq:Brel3}), where we have additionally identified $q_k=2p+|m|+1$. 
\begin{table}[h!]
\centering
\caption{\bf Coefficients $\gamma_{mnrl}$ for $M=4$ lasing modes.}
\begin{tabular}{cccc}
\hline
coefficient & expression & coefficient & expression \\
\hline
 $\gamma_{1111}$ & $1/\pi$ &  $\gamma_{1444}$ & $-1/(6\sqrt{3}\pi)$ \\
 $\gamma_{1114}$ & $1/(2\sqrt{3}\pi)$ &  $\gamma_{2222}$ & $3/(4\pi)$\\
 $\gamma_{1122}$ & $1/(2\pi)$ &  $\gamma_{2233}$ & $1/(4\pi)$\\
 $\gamma_{1133}$ & $1/(2\pi)$ &  $\gamma_{2244}$ & $(4+\sqrt{2})/(12\pi)$\\
 $\gamma_{1144}$ & $1/(3\pi)$ &  $\gamma_{3333}$ & $3/(4\pi)$\\
 $\gamma_{1224}$ & $-1/(2\sqrt{6}\pi)$ &  $\gamma_{3344}$ & $(4-\sqrt{2})/(12\pi)$\\
 $\gamma_{1334}$ & $1/(2\sqrt{6}\pi)$ &  $\gamma_{4444}$ & $23/(36\pi)$\\
\hline
\end{tabular}
  \label{tab:gamma}
\end{table}
\section{Coefficients $\gamma_{mnrl}$ and beating frequencies}\label{app:coeff}
As highlighted in the main text, the number of coefficients $\gamma_{mnrl}$ different from each other for the entire set of equations is given by the binomial coefficient $\binom{M+4-1}{4}$ with $M$ being the number of modes relevant for the laser dynamics. Therefore, with $M=4$, as in our case, the number of $\gamma_{mnrl}$ is 35. However, given the symmetry (and anti-symmetry) of functions $C_1$, $C_2$, $C_3$, $C_4$ with respect to the Cartesian coordinates $x$ and $y$, only 14 of these coefficient are non-null. We report such non-null coefficients $\gamma_{mnrl}$ in Table \ref{tab:gamma}.
Note that the coefficients in Table \ref{tab:gamma} account also for all permutations of the displayed indices (e.g., $\gamma_{1114}=\gamma_{4111}=\gamma_{1411}=\gamma_{1141}$). While such permutations lead to the same coefficient $\gamma_{mnrl}$, they can result in different beating frequencies. We report in Table \ref{tab:beatings} all the possible angular beating frequencies resulting from each one of the coefficient in Table \ref{tab:gamma}.\\

\begin{table}[h!]
\centering
\caption{\bf Beating frequencies arising from each $\gamma_{mnrl}$ in Table \ref{tab:gamma} and related permutations (with repetition) for $M=4$.}
\begin{tabular}{cc}
\hline
coefficient & angular beating frequencies \\
\hline
$\gamma_{1111}$ & $0$ \\
$\gamma_{1114}$ & $\omega_4$, $-\omega_4$\\
$\gamma_{1122}$ & 0, $2\omega_2$, $-2\omega_2$\\
$\gamma_{1133}$ & 0, $2\omega_3$, $-2\omega_3$\\
$\gamma_{1144}$ & 0, $2\omega_4$, $-2\omega_4$\\
$\gamma_{1224}$ & $\omega_4$, $-\omega_4$, $\omega_4-2\omega_2$, $-(\omega_4-2\omega_2)$\\
$\gamma_{1334}$ & $\omega_4$, $-\omega_4$, $\omega_4-2\omega_3$, $-(\omega_4-2\omega_3)$\\
$\gamma_{1444}$ & $\omega_4$, $-\omega_4$\\
$\gamma_{2222}$ & $0$\\
$\gamma_{2233}$ & 0, $2(\omega_3-\omega_2)$, $-2(\omega_3-\omega_2)$\\
$\gamma_{2244}$ & 0, $2(\omega_4-\omega_2)$, $-2(\omega_4-\omega_2)$\\
$\gamma_{3333}$ & 0\\
$\gamma_{3344}$ & 0, $2(\omega_4-\omega_3)$, $-2(\omega_4-\omega_3)$\\
$\gamma_{4444}$ & 0\\
\hline
\end{tabular}
  \label{tab:beatings}
\end{table}
When instead the number of lasing modes is $M=5$ then only 22 of the 70 possible $\gamma_{mnrl}$ are not null. Additional coefficients and related beating frequencies, to be added to those of Tables \ref{tab:gamma} and \ref{tab:beatings}, respectively, are reported in Tables \ref{tab:M5gamma} and \ref{tab:M5beatings}.
\begin{table}[ht!]
\centering
\caption{\bf Additional coefficients $\gamma_{mnrl}$ for $M=5$ lasing modes.}
\begin{tabular}{cccc}
\hline
coefficient & expression & coefficient & expression \\
\hline
 $\gamma_{1155}$ & $1/(4\pi)$ &  $\gamma_{2345}$ & $-1/(8\sqrt{3}\pi)$\\
 $\gamma_{1235}$ & $1/(4\pi)$ & $\gamma_{3355}$ & $3/(8\pi)$\\
 $\gamma_{1455}$ & $-1/(8\sqrt{3}\pi)$ & $\gamma_{4455}$ & $5/(24\pi)$ \\
 $\gamma_{2255}$ & $3/(8\pi)$ &  $\gamma_{5555}$ & $9/(16\pi)$\\ 
\hline
\end{tabular}
  \label{tab:M5gamma}
\end{table}

\begin{table}[ht!]
\centering
\caption{\bf Additional beating frequencies arising from each $\gamma_{mnrl}$ in Table \ref{tab:M5gamma} for $M=5$.}
\begin{tabular}{cc}
\hline
coefficient & angular beating frequencies \\
\hline
 $\gamma_{1155}$ &  $0$, $\pm2\omega_5$ \\
 $\gamma_{1235}$ & $\pm(\omega_2+\omega_5-\omega_3)$, $\pm(\omega_2+\omega_3-\omega_5)$,\\
&  $\pm(\omega_3+\omega_5-\omega_2)$\\
 $\gamma_{1455}$ & $\pm\omega_4$, $\pm(\omega_4-2\omega_5)$ \\
 $\gamma_{2255}$ &  $0$, $\pm2(\omega_2-\omega_5)$\\
 $\gamma_{2345}$ & $\pm(\omega_3-\omega_2+\omega_5-\omega_4)$, $\pm(\omega_3-\omega_2+\omega_4-\omega_5)$,\\
 & $\pm(\omega_4-\omega_2+\omega_5-\omega_3)$\\
 $\gamma_{3355}$ &   $0$, $\pm2(\omega_3-\omega_5)$\\
 $\gamma_{4455}$ &  $0$, $\pm2(\omega_4-\omega_5)$\\
$\gamma_{5555}$ &  0\\
\hline
\end{tabular}
  \label{tab:M5beatings}
\end{table}

\section{Only beating frequencies emerging while not properly accounting coherent frequency mixing}\label{app:incorrect_mixing}
As discussed in the main text, if instead of $|E(\rho,\phi,t)|^2=|\sum_{m=1}^4E_m(t)C_m(\rho,\phi)|^2$, we considered  $\sum_{m=1}^4|E_m(t)C_m(\rho,\phi)|^2$ (as e.g. in \cite{Law1997}) in the approximation for $f_m(t)e^{-i\omega_m t}$ in Section \ref{sec:ellipt}, this would result in the following expression for $f_m(t)e^{-i\omega_m t}$
\begin{eqnarray*}
&\hspace{-73mm}f_m(t)e^{-i\omega_mt}=\\
&\hspace{-3mm}=\Lambda \int_0^{\infty}\int_0^{2\pi}\hspace{-1mm}\rho \D\rho\D\phi \sum_n\hspace{-0.5mm}\tilde{E}_n(t)C_nC_m e^{i\omega_{nm}t}\hspace{-1mm}\left[1-\sum_l\hspace{-0.5mm}\frac{|E_l(t)|^2}{E_s^2}C_l^2\right]\\
&\hspace{-53mm}=\Lambda \left[\sum_n \delta_{mn}\tilde{E}_n(t)e^{i\omega_{nm}t}\right.\\
&\left.-\sum_{n,l}\int_0^{\infty}\int_0^{2\pi}\rho \D\rho\D\phi C_n C_m C_l^2\frac{\tilde{E}_n(t)|E_l(t)|^2}{E_s^2}e^{i\omega_{nm}t}\right]\\
&\hspace{-25mm}=\Lambda\left[\tilde{E}_m(t)-\sum_{n,l} \gamma_{mnll}\frac{\tilde{E}_n(t)|\tilde{E}_l(t)|^2}{E_s^2}e^{i\omega_{nm}t}\right]
\end{eqnarray*}
In this case, the only beating frequencies emerging from the terms with non-null $\gamma_{mnll}$ coefficients are $\omega_4$, $-\omega_4$, resulting from the terms with $\gamma_{1411}$, $\gamma_{1422}$, $\gamma_{1433}$, $\gamma_{1444}$ and permutations of the first two indices.\\
As an example, we show in the following  how this choice impacts the differential equation for the total power from the VCSEL $P_{tot}(t)=\sum_m|E_m(t)|^2$, depending on the considered lasing modes.
\begin{eqnarray*}
&\hspace{-30mm}\frac{\D P_{tot}(t)}{\D t}=\sum_{m}\left[ \tilde{E}_m(t) \frac{\D \tilde{E}_m^*(t)}{\D t}+\tilde{E}_m^*(t) \frac{\D \tilde{E}_m(t)}{\D t}\right]\\
&\hspace{-42mm}=\sum_{m}\left\{\left(-\frac{1}{\tau_{p,m}}+\Gamma G_N\Lambda\right)|\tilde{E}_m(t)|^2\right.\\
&\hspace{-5mm}\left.-\frac{\Gamma G_N\Lambda}{E_s^2}\hspace{-0.5mm}\sum_{n,l}\hspace{-0.5mm}\gamma_{mnll}|\tilde{E}_l(t)|^2|\tilde{E}_m^*(t)\tilde{E}_n(t)|\hspace{-0.5mm}\left[\cos{\zeta(t)}+\alpha \sin{\zeta(t)}\right]\hspace{-0.5mm}\right\}\\
\end{eqnarray*}
with $\zeta(t)=\phi_n(t)-\phi_m(t)+\omega_{nm}t$.\\
If the only modes to be lasing are $C_1$ and $C_2$, then 
$\gamma_{1111}$, $\gamma_{1122}$, $\gamma_{1122}$, $\gamma_{2222}$ (and permutations of the first two indices) will be the only non-null coefficients in the sum over $(m,n,l)$, because all other indices combinations lead to odd functions in at least one integration variable with a domain that is symmetric with respect to the origin. It can then be easily verified that none of these terms gives rise to a beating frequency.\\
If instead the only modes to to be lasing are $C_1$ and $C_4$, then $\gamma_{1111}$, $\gamma_{1144}$, $\gamma_{1411}$, $\gamma_{1444}$ (and permutations of the first two indices) are the only non-null coefficients in the sum. Of these $\gamma_{1411}$ and $\gamma_{1444}$ lead to a beating frequency at $\omega_4$ (and $\gamma_{4111}$ and $\gamma_{4144}$ to a beating frequency at $-\omega_4$), which results in the following equation for $P_{tot}(t)$
\begin{eqnarray*}
&\hspace{-23mm}\frac{\D P_{tot}(t)}{\D t}=\sum_{m=1,4}\left\{\left(-\frac{1}{\tau_{p,m}}+\Gamma G_N\Lambda\right)|\tilde{E}_m(t)|^2\right.\\
&\left.-\frac{\Gamma G_N\Lambda}{E_s^2}\left[\gamma_{mmmm}|E_m(t)|^4\hspace{-0.5mm}+\hspace{-0.5mm}\gamma_{mmnn_{m\neq n}}|E_m(t)|^2|E_n(t)|^2\right]\hspace{-0.5mm}\right\}\\
&\hspace{-32mm}-2\frac{\Gamma G_N\Lambda}{E_s^2}|\tilde{E}_1^*(t)\tilde{E}_4(t)|\left[\gamma_{1411}|E_1(t)|^2\right.\\
&\hspace{-26mm}\left.+\gamma_{1444}|E_4(t)|^2\right]\cos{\left[\phi_4(t)-\phi_1(t)+\omega_4 t\right]}
\end{eqnarray*}
We highlight that two conditions are necessary to observe an effect of this beating frequency in the total power dynamics: (i) to have a receiver bandwidth large enough to capture the frequency beating, (ii) to be considering an aperture geometry that allows for the lasing of at least two modes $C_m$, $C_n$ such that $C_m(x,y) C_n(x,y)$ is an even function both in $x$ and $y$. Finally, we highlight that, in other literature approaches such as \cite{Valle1995}, where only the photon number is dynamically described, any beating is impossible to observe a priori.
}

% Bibliography
\bibliography{biblio}

\begin{thebibliography}{10}
\providecommand{\url}[1]{#1}
\csname url@samestyle\endcsname
\providecommand{\newblock}{\relax}
\providecommand{\bibinfo}[2]{#2}
\providecommand{\BIBentrySTDinterwordspacing}{\spaceskip=0pt\relax}
\providecommand{\BIBentryALTinterwordstretchfactor}{4}
\providecommand{\BIBentryALTinterwordspacing}{\spaceskip=\fontdimen2\font plus
\BIBentryALTinterwordstretchfactor\fontdimen3\font minus \fontdimen4\font\relax}
\providecommand{\BIBforeignlanguage}[2]{{%
\expandafter\ifx\csname l@#1\endcsname\relax
\typeout{** WARNING: IEEEtran.bst: No hyphenation pattern has been}%
\typeout{** loaded for the language `#1'. Using the pattern for}%
\typeout{** the default language instead.}%
\else
\language=\csname l@#1\endcsname
\fi
#2}}
\providecommand{\BIBdecl}{\relax}
\BIBdecl

\bibitem{Bhatt2024}
V.~Bhatt, ``Multimode links based on high-speed vcsels for cost-effective data center connectivity,'' in \emph{Optical Fiber Communication Conference (OFC) 2024}.\hskip 1em plus 0.5em minus 0.4em\relax Optica Publishing Group, 2024, p. Th1B.4.

\bibitem{Gazula2019}
D.~Gazula, N.~Chitica, M.~Chacinski, G.~Landry, and J.~Tatum, ``{VCSEL} with elliptical aperture having reduced {RIN},'' U.S. Patent 20190341743A1, 2019.

\bibitem{Wang2019}
B.~Wang, W.~Sorin, M.~Tan, S.~Mathai, and S.~Cheung, ``Intensity noise mitigation for vertical-cavity surface emitting lasers,'' U.S. Patent 10985531B2, 2019.

\bibitem{Debernardi2002}
P.~Debernardi, G.~Bava, C.~Degen, I.~Fischer, and W.~Elsasser, ``Influence of anisotropies on transverse modes in oxide-confined {VCSEL}s,'' \emph{IEEE Journal of Quantum Electronics}, vol.~38, no.~1, pp. 73--84, 2002.

\bibitem{Lenstra2014}
\BIBentryALTinterwordspacing
D.~Lenstra and M.~Yousefi, ``Rate-equation model for multi-mode semiconductor lasers with spatial hole burning,'' \emph{Opt. Express}, vol.~22, no.~7, pp. 8143--8149, Apr 2014. [Online]. Available: \url{https://opg.optica.org/oe/abstract.cfm?URI=oe-22-7-8143}
\BIBentrySTDinterwordspacing

\bibitem{Furfaro2004}
L.~Furfaro, F.~Pedaci, M.~Giudici, X.~Hachair, J.~Tredicce, and S.~Balle, ``Mode-switching in semiconductor lasers,'' \emph{IEEE Journal of Quantum Electronics}, vol.~40, no.~10, pp. 1365--1376, 2004.

\bibitem{Chacinski2010}
M.~Chaci\'{n}ski and R.~Schatz, ``Impact of losses in the bragg section on the dynamics of detuned loaded {DBR} lasers,'' \emph{IEEE Journal of Quantum Electronics}, vol.~46, no.~9, pp. 1360--1367, 2010.

\bibitem{Bardella2017}
\BIBentryALTinterwordspacing
P.~Bardella, L.~L. Columbo, and M.~Gioannini, ``Self-generation of optical frequency comb in single section quantum dot {F}abry-{P}erot lasers: a theoretical study,'' \emph{Opt. Express}, vol.~25, no.~21, pp. 26\,234--26\,252, Oct 2017. [Online]. Available: \url{https://opg.optica.org/oe/abstract.cfm?URI=oe-25-21-26234}
\BIBentrySTDinterwordspacing

\bibitem{Prati1994_chaos}
\BIBentryALTinterwordspacing
F.~Prati, A.~Tesei, L.~Lugiato, and R.~Horowicz, ``Stable states in surface-emitting semiconductor lasers,'' \emph{Chaos, Solitons \& Fractals}, vol.~4, no.~8, pp. 1637--1654, 1994, special Issue: Nonlinear Optical Structures, Patterns, Chaos. [Online]. Available: \url{https://www.sciencedirect.com/science/article/pii/0960077994901015}
\BIBentrySTDinterwordspacing

\bibitem{Tatum2015}
J.~A. Tatum, D.~Gazula, L.~A. Graham, J.~K. Guenter, R.~H. Johnson, J.~King, C.~Kocot, G.~D. Landry, I.~Lyubomirsky, A.~N. MacInnes, E.~M. Shaw, K.~Balemarthy, R.~Shubochkin, D.~Vaidya, M.~Yan, and F.~Tang, ``{VCSEL}-based interconnects for current and future data centers,'' \emph{Journal of Lightwave Technology}, vol.~33, no.~4, pp. 727--732, 2015.

\bibitem{Quirce2011}
\BIBentryALTinterwordspacing
A.~Quirce, A.~Valle, C.~Gim\'{e}nez, and L.~Pesquera, ``Intensity noise characteristics of multimode {VCSEL}s,'' \emph{J. Lightwave Technol.}, vol.~29, no.~7, pp. 1039--1045, Apr 2011. [Online]. Available: \url{https://opg.optica.org/jlt/abstract.cfm?URI=jlt-29-7-1039}
\BIBentrySTDinterwordspacing

\bibitem{Michalzik2013}
\BIBentryALTinterwordspacing
R.~Michalzik, \emph{VCSELs}.\hskip 1em plus 0.5em minus 0.4em\relax Berlin, Heidelberg: Springer Berlin Heidelberg, 2013. [Online]. Available: \url{https://doi.org/10.1007/978-3-642-24986-0}
\BIBentrySTDinterwordspacing

\bibitem{Prati1994}
F.~Prati, M.~Brambilla, and L.~Lugiato, ``Pattern formation in lasers,'' \emph{Riv. Nuovo Cim.}, vol.~17, no.~3, pp. 1--85, 1994.

\bibitem{Rimoldi2022}
\BIBentryALTinterwordspacing
C.~Rimoldi, L.~L. Columbo, J.~Bovington, S.~Romero-Garc\'{i}a, and M.~Gioannini, ``Damping of relaxation oscillations, photon-photon resonance, and tolerance to external optical feedback of {III}-{V}/{S}i{N} hybrid lasers with a dispersive narrow band mirror,'' \emph{Opt. Express}, vol.~30, no.~7, pp. 11\,090--11\,109, Mar 2022. [Online]. Available: \url{http://opg.optica.org/oe/abstract.cfm?URI=oe-30-7-11090}
\BIBentrySTDinterwordspacing

\bibitem{Valle1995}
A.~Valle, J.~Sarma, and K.~Shore, ``Spatial holeburning effects on the dynamics of vertical cavity surface-emitting laser diodes,'' \emph{IEEE Journal of Quantum Electronics}, vol.~31, no.~8, pp. 1423--1431, 1995.

\bibitem{Coldren}
L.~A. Coldren and S.~W. Corzine, \emph{Semiconductor and Photonic Integrated Circuits}.\hskip 1em plus 0.5em minus 0.4em\relax John Wiley \& Sons, Ltd, 1995.

\bibitem{Gao2012}
\BIBentryALTinterwordspacing
J.~Gao, ``High frequency modeling and parameter extraction for vertical-cavity surface emitting lasers,'' \emph{J. Lightwave Technol.}, vol.~30, no.~11, pp. 1757--1763, Jun 2012. [Online]. Available: \url{https://opg.optica.org/jlt/abstract.cfm?URI=jlt-30-11-1757}
\BIBentrySTDinterwordspacing

\bibitem{Hamad2020}
W.~Hamad, M.~Bou~Sanayeh, M.~M. Hamad, and W.~H.~E. Hofmann, ``Impedance characteristics and chip-parasitics extraction of high-performance {VCSEL}s,'' \emph{IEEE Journal of Quantum Electronics}, vol.~56, no.~1, pp. 1--11, 2020.

\bibitem{Hamed2019}
W.~Hamad, M.~B. Sanayeh, T.~Siepelmeyer, H.~Hamad, and W.~H.~E. Hofmann, ``Small-signal analysis of high-performance {VCSEL}s,'' \emph{IEEE Photonics Journal}, vol.~11, no.~2, pp. 1--12, 2019.

\bibitem{Minelli2024}
L.~Minelli, F.~Forghieri, T.~Shao, A.~Shahpari, and R.~Gaudino, ``Tdecq-based optimization of nonlinear digital pre-distorters for {VCSEL}-{MMF} optical links using end-to-end learning,'' \emph{Journal of Lightwave Technology}, vol.~42, no.~2, pp. 621--635, 2024.

\bibitem{Gil2014}
\BIBentryALTinterwordspacing
L.~Gil and G.~L. Lippi, ``Phase instability in semiconductor lasers,'' \emph{Phys. Rev. Lett.}, vol. 113, p. 213902, Nov 2014. [Online]. Available: \url{https://link.aps.org/doi/10.1103/PhysRevLett.113.213902}
\BIBentrySTDinterwordspacing

\bibitem{Law1997}
J.~Law and G.~Agrawal, ``Mode-partition noise in vertical-cavity surface-emitting lasers,'' \emph{IEEE Photonics Technology Letters}, vol.~9, no.~4, pp. 437--439, 1997.

\bibitem{Bava2001}
\BIBentryALTinterwordspacing
G.~P. Bava, P.~Debernardi, and L.~Fratta, ``Three-dimensional model for vectorial fields in vertical-cavity surface-emitting lasers,'' \emph{Phys. Rev. A}, vol.~63, p. 023816, Jan 2001. [Online]. Available: \url{https://link.aps.org/doi/10.1103/PhysRevA.63.023816}
\BIBentrySTDinterwordspacing

\bibitem{Debernardi2003}
P.~Debernardi and G.~P. Bava, ``Coupled mode theory: a powerful tool for analyzing complex {VCSEL}s and designing advanced device features,'' \emph{IEEE Journal of Selected Topics in Quantum Electronics}, vol.~9, no.~3, pp. 905--917, 2003.

\bibitem{Moser2015}
P.~Moser, J.~A. Lott, G.~Larisch, and D.~Bimberg, ``Impact of the oxide-aperture diameter on the energy efficiency, bandwidth, and temperature stability of 980-nm {VCSEL}s,'' \emph{Journal of Lightwave Technology}, vol.~33, no.~4, pp. 825--831, 2015.

\end{thebibliography}
\bibliographystyle{IEEEtran}
\begin{IEEEbiographynophoto}{Cristina Rimoldi}
received her BSc and MSc degrees in Physics at the Universit\`a degli Studi dell'Insubria (Como, Italy) in 2011 and 2014, respectively. She obtained her PhD diploma at the Universit\'e C\^ote d'Azur (Nice, France) in 2017 with a thesis on extreme events in nonlinear optical cavities. From 2018 to 2020 she was a Post-Doc at the Institut National de la Recherche Scientifique-Énergie Matériaux Télécommunications (INRS-EMT) in Varennes (Quebec, Canada), where she worked on the modeling of optical cavities comprising integrated photonic components. Since October 2020, she joined the Dipartimento di Elettronica e Telecomunicazioni at Politecnico di Torino (Torino, Italy). Her current research is focused on the theoretical and numerical study of hybrid lasers in silicon photonic platforms. Other research interests include multimode VCSELs, laser nonlinear dynamics, optical rogue waves, and cavity solitons.\vspace{-22pt}\\
\end{IEEEbiographynophoto}
\begin{IEEEbiographynophoto}{Lorenzo L. Columbo}
received the MSc degree in Physics from the Universit\`a degli Studi di Bari (Bari, Italy) in 2002 and the PhD in Physics and Astrophysics from the Dipartimento di Scienza e Alta Tecnologia of the
Universit\`a degli Studi dell’Insubria (Como, Italy) in 2007. He worked as
reasercher in the field of self-organization phenomena in semiconductor lasers
at the Institut Nonlin\'{e}aire de Nice (Valbonne, France), at the Dipartimento
di Fisica of the Universit\`a degli Studi di Bari (Bari, Italy) and at the
Dipartimento di Scienza e Alta Tecnologia of the Universit\`a degli Studi
dell’Insubria (Como, Italy). In April 2016 he joined the Dipartimento di
Elettronica e Telecomunicazioni of the Politecnico di Torino (Torino, Italy), where he is currently Associate Professor.
Here, his theoretical research activity focuses on the modeling of 1. optical frequency combs and solitons self-generation in Quantum Dots lasers and Quantum Cascade lasers and 2.
hybrid laser dynamics in silicon photonics platform. His main fields of interest are:
Quantum Electronics, Optoelectronics and Nonlinear Optics.\vspace{-50pt}\\
\end{IEEEbiographynophoto}
\begin{IEEEbiographynophoto}{Alberto Tibaldi}
Alberto Tibaldi received the B.Sc., M.Sc., and Ph.D. degrees in electronic engineering from the Polytechnic of Turin, Italy, in 2009, 2011, and 2015, respectively. From 2012 to 2019, he was with the Italian National Council of Research (CNR) as a Research Fellow. Since 2019, he has been with the Department of Electronics and Telecommunications, Polytechnic of Turin as an Assistant Professor, where he teaches courses on semiconductor devices and numerical analysis. His scientific research interests focuses on the multi-physics modeling of optoelectronic devices.\vspace{-50pt}\\
\end{IEEEbiographynophoto}
\begin{IEEEbiographynophoto}{Pierluigi Debernardi}
Pierluigi Debernardi was born in Casale Monferrato, Italy. He received a degree in electronics engineering from Politecnico di Torino, Turin, Italy, in 1987. Since 1989, he has been with the Italian National Council of Research, Politecnico di Torino. His interests include the modeling of semiconductor materials and devices for optoelectronic applications. He is mostly involved in modeling and designing VCSEL structures with noncircular and/or complex geometries, so as to achieve specific performances.\vspace{-50pt}\\
\end{IEEEbiographynophoto}
\begin{IEEEbiographynophoto}{Sebastian Romero Garc\'ia}
Sebastian Romero-Garcia was born in Rute, Spain, in 1984. He received the M.Sc. degree in telecommunications engineering from the University of Malaga, Spain, in 2009 and the Dr.-Ing. degree from RWTH Aachen University, Germany, in 2016. From 2009 to 2012, he was working as a Research Assistant with the Photonics \& RF Research Lab, University of Malaga. In 2012, he joined the Institute of Integrated Photonics at RWTH Aachen where he worked on the development of high-performance components for silicon photonics aimed at datacom applications as well as silicon nitride photonic integrated circuits for biosensing. Since 2017 he is a Hardware Engineer at Cisco working in the research and development of high-speed transceivers for next-generation optical communication systems.\vspace{-50pt}\\
\end{IEEEbiographynophoto}
\begin{IEEEbiographynophoto}{Christian Raabe}
has an Electrical Engineering degree from Braunschweig Technical University (Germany). Since 2010, he is Technical Leader Engineer at Cisco Optical GmbH, where he focuses on the design of high speed optical interfaces for digital optical communication systems.\vspace{-50pt}\\
\end{IEEEbiographynophoto}
\IEEEpubidadjcol
\begin{IEEEbiographynophoto}{Mariangela Gioannini}
received the MS degree in Electronic Engineering and the PhD degree Electronics and Communication Engineering both from Politecnico di Torino in 1998 and 2002 respectively. Since 2005 she is with Dipartimento di Elettronica e Telecomunicazioni of Politecnico di Torino where she is now full professor. She has carried on research on the numerical modelling of semiconductor lasers and optical amplifiers based on III-V materials including quantum dot lasers and QCLs. In recent years research activity has been extended to silicon photonics and VCSELs including the coordination of an experimental lab dedicated to VCSELs and silicon photonics PICs. Since 2013 she has been in the technical program committee of major conferences in the area of photonics: CLEO-Europe, CLEO US, IEEE ISLC and Photonics West. Since 2022 she is Associate Editor of IEEE Photonic Technology Letters.
\end{IEEEbiographynophoto}

\end{document}